\documentclass[preprint,5p,times,authoryear]{elsarticle}
\usepackage{amssymb}
\usepackage{amsmath}
\usepackage{booktabs}
\usepackage{graphicx}
\usepackage{subcaption}
\usepackage{tabularx}
\usepackage{array} 
\usepackage{xcolor}
\usepackage{threeparttable}
\usepackage{afterpage}
\newcolumntype{Y}{>{\centering\arraybackslash}X}

\usepackage[colorlinks=true, linkcolor=blue, citecolor=blue, urlcolor=blue]{hyperref}

\makeatletter
\def\cormark#1{}   
\def\fnmark#1{}    
\makeatother

\begin{document}

\begin{frontmatter}

\title{FusionVul: A Multimodal Feature Fusion Framework for Source Code Vulnerability Detection}

\author[a,b]{Hongyu Yang}
\ead{yhyxlx@hotmail.com} 

\author[a]{Yaping Zhu\corref{cor1}}
\ead{ypzhu111@163.com}

\author[a]{Jingchuan Luo}
\ead{kafu2001@163.com}

\author[c]{Hiroshi Nomaguchi}
\ead{d8211102@u-aizu.ac.jp}

\author[c]{Chunhua Su}
\ead{chsu@u-aizu.ac.jp}

\author[d]{Willy Susilo}
\ead{wsusilo@uow.edu.au}

\cortext[cor1]{Corresponding author.}

\affiliation[a]{organization={School of Safety Science and Engineering, Civil Aviation University of China},
            city={Tianjin},
            postcode={300300},
            country={China}}
\affiliation[b]{organization={School of Computer Science and Technology, Civil Aviation University of China},
            city={Tianjin},
            postcode={300300}, 
            country={China}}
\affiliation[c]{organization={School of Computer Science and Engineering, University of Aizu},
            city={Aizu-Wakamatsu},
            postcode={965-8580}, 
            state={Fukushima},
            country={Japan}}
\affiliation[d]{organization={School of Computing and Information Technology, University of Wollongong},
            city={Wollongong},
            postcode={2522}, 
            state={NSW},
            country={Australia}}

\begin{abstract}
Source code vulnerability detection remains a long-standing challenge due to the increasing scale, structural complexity, and semantic diversity of modern codebases. Conventional static-analysis or rule-based approaches often fail to capture subtle execution dependencies, while single-modality learning models tend to overlook critical structural information embedded beyond the lexical surface of source code. 
To improve robustness across heterogeneous code patterns, we propose FusionVul, a joint representation learning framework that integrates sequential syntactic representations extracted by a pretrained Transformer encoder with structural semantics propagated through a graph neural network. The framework further incorporates a cross-attention-based feature fusion network to enable fine-grained cross-modal interaction and employs a sample-aware weighting mechanism to integrate multiple predictive branches. 
Experimental results on four datasets demonstrate that FusionVul achieves superior F1 scores on datasets with highly dispersed function size distributions and broader vulnerability-type coverage, such as SVulD and DiverseVul, reflecting its capability to capture complex and diverse vulnerability patterns. 

\end{abstract}




\end{frontmatter}



\section{Introduction}\label{sec1}

Security-critical flaws embedded in software source code remain a dominant root cause of system vulnerabilities. As contemporary software continues to expand in scale, complexity, and structural diversity, the range of possible defect patterns has similarly broadened. This evolution renders the reliable identification of latent vulnerabilities a persistent challenge in both cybersecurity and software engineering research. Over the past decade, numerous defect-detection techniques have been developed to address this need ~\cite{1Diversevul,2,3,5}. Although traditional rule-based static analysis tools and manually crafted detection heuristics retain value in specific settings, they typically depend on intensive expert effort, impose substantial maintenance overhead, and tend to miss concealed execution paths or generate spurious alarms when confronted with deeply nested or implicit control structures ~\cite{12Knighter}. Dynamic fuzzing approaches ~\cite{nourry2025my}, despite their ability to uncover hidden runtime behaviors, are inherently constrained by execution environments and the size of the input search space, which restricts their coverage in large-scale systems.

Recently, advances in large-scale pretrained code models ~\cite{7Openaicodex,8Codellama,9DeepSeekCoder,10Finetuning} and graph neural networks (GNNs) ~\cite{15GGNN,13GGCNN} have introduced promising opportunities for automating vulnerability detection. Some approaches ~\cite{16Unixcoder,34} exploit pretrained code models to capture linear textual patterns from source code, whereas others ~\cite{26DeepReveal} emphasize graph-structured representations such as the Code Property Graph (CPG) to model data- and control-flow dependencies. 
However, source code can be represented in a multimodal manner as both a linear sequence and a property graph, and these modalities often provide complementary semantic cues. 
Consequently, reliance on a single modality often leads to incomplete semantic understanding and weakens the ability to characterize complex program behaviors. Existing multimodal methods exhibit notable limitations in feature fusion and utilization. 
Although UniXcoder embeds Abstract Syntax
Tree (AST) information into token representations during pre-training, this fusion is implicit and models only surface-level syntactic information.
Furthermore, some approaches perform implicit fusion at the property-graph node embedding stage ~\cite{CSGVD,Linevd}, still without explicitly modeling fine-grained semantic correspondences between sequence and structural features, which constrains the expressiveness of the fused representation. 
Moreover, datasets differ in vulnerability types and code-structure complexity (characterized and quantified in Section 4.2); thus, the degree of reliance on sequential versus structural features can vary at the final decision stage. For example, vulnerabilities involving unsafe API invocations often exhibit salient signals in token-level sequential semantics, whereas vulnerabilities related to improper data flows rely more heavily on the data-flow structures captured by the CPG, and more complex vulnerabilities may further require complementary information from cross-modal fusion. However, many existing methods aggregate independent sequential and structural branches with a fixed weighting strategy at inference time~\cite{19comprehensivecodegraphs,22Vul-LMGNNs}. This limits the effective use of cross-modal fused representations and makes it difficult to fully capture the complementary roles of different semantic branches during decision-making. To some extent, this restricts the sufficiency of feature utilization and the generalization ability of the model.

To address these challenges, we introduce FusionVul, a multimodal feature fusion framework for code vulnerability detection that emphasizes explicit cross-modal feature interactions and weighted aggregation across multiple predictive branches. FusionVul consists of three modules: a modality-specific feature extraction module, a cross-attention-based feature fusion module, and a multi-branch weighted aggregation module. 
Specifically, at the feature extraction stage, 
the pre-trained UniXcoder model ~\cite{16Unixcoder}, built on a Transformer architecture, is used to derive sequential syntactic representations from the Linear Code Sequence (LCS), 
while a Gated Graph Neural Network (GGNN) aggregates behavioral semantics from the CPG. This hybrid feature extraction design enables the model to capture complementary semantics from both sequential and graph-structured perspectives. At the feature fusion stage, to promote deep semantic interactions between sequential and structural features, we draw inspiration from recent cross-attention research ~\cite{21CrossFuse}, and further develop a Cross-Attention Feature Fusion Network (CAFFNet).
CAFFNet allows LCS tokens to attend to relevant structural representations derived from the CPG, facilitating fine-grained multimodal semantic fusion and enriching the fused representations. Finally, at the prediction stage, considering that different vulnerability patterns may emphasize semantic cues from different modalities, we propose a Sample-Aware Weighting (SaW) strategy.
Built upon existing dual-branch designs~\cite{22Vul-LMGNNs}, this strategy further incorporates the fused features and performs weighted aggregation over the predictions from three branches, namely the sequential, structural, and fused branches, at the decision stage. This strategy preserves decision signals from individual modalities, while exploiting the complementary information provided by the fused branch, thereby improving feature utilization and enhancing the model’s generalization capability across datasets with diverse characteristics.

Overall, by leveraging CAFFNet to facilitate deep cross-modal feature fusion and employing a SaW strategy at the decision stage to enhance the utilization of multi-branch features, FusionVul aims to alleviate the limitations of cross-modal semantic fusion and the insufficient exploitation of fused representations in vulnerability detection.
The implementation details and dataset information are available at: https://github.com/Julia-ypz/FusionVul.

The key contributions of this study are summarized as follows:
\begin{itemize}
\item We propose FusionVul, a multimodal vulnerability detection framework that emphasizes feature fusion and utilization, in which UniXcoder captures sequential-syntactic representations from the LCS, while a GGNN encodes structural-semantic features from the CPG. This dual-view design enables complementary code representations.
\item We design CAFFNet, which models fine-grained semantic correlations between LCS features and CPG features via a cross-attention mechanism, thereby enhancing the expressiveness of the fused representations. 
\item We further propose a SaW strategy to combine prediction outputs from the LCS, CPG, and fusion branches. This strategy integrates information from multiple semantic perspectives at the prediction stage, enhancing the effective utilization of features. 
\end{itemize}
\section{Related Work}\label{sec2}
\subsection{Code Representation and Detection Based on LCS}
In recent years, large-scale pre-trained models for source code understanding have gained increasing prominence in automated software analysis ~\cite{23Codebert,24Graphcodebert,35Longcoder}. These models incorporate semantic cues such as data-flow relations and signals derived from the AST during the pre-training stage, which enables substantial advancements in downstream applications including code retrieval, summarization, and vulnerability detection. For example, GraphCodeBERT ~\cite{24Graphcodebert} enriches token embeddings by injecting data flow-aware semantic associations among variables, whereas UniXcoder ~\cite{16Unixcoder} employs a joint pre-training paradigm that integrates AST linearization and natural language comments, offering flexible encoder-decoder configurations. These sequence-oriented pre-trained encoders have thus become fundamental components of vulnerability detection frameworks relying on LCS or comment-enhanced textual inputs. Nonetheless, sequence-only approaches can struggle to capture deeper logical flaws or long-range structural dependencies, particularly in code fragments that appear simple textually yet exhibit complex underlying structure.

\subsection{Code Representation and Detection Based on CPG}
The CPG explicitly models structural and dependency relationships that token-based sequential representations fail to surface. A wide range of GNNs ~\cite{19comprehensivecodegraphs,13GGCNN,15GGNN} has demonstrated substantial capability in extracting such graph-oriented information ~\cite{17Devign,26DeepReveal,27Regvd}. Among them, the GGNN ~\cite{15GGNN} employs gated message propagation to perform multi-hop information aggregation, thereby yielding expressive node-level representations. Prior studies leveraging CPG-rich graph structures ~\cite{17Devign,26DeepReveal} indicate that GNN-based methods excel in tasks such as vulnerability localization and defect identification. Consequently, CPG representations and LCS encodings are widely recognized as complementary modalities within modern vulnerability detection systems.

\subsection{Cross-Modal Feature Fusion}
Given the inherent semantic divergence between textual sequences and graph-structured code representations, various efforts have explored the integration of LCS and CPG modalities.
In this work, we use the term \textit{hybrid system} to refer to approaches that combine a pretrained Transformer-based language model (as a sequential encoder) with a graph neural network (as a structural encoder).   
Several hybrid systems ~\cite{Linevd,31,32} explore the technical feasibility of jointly optimizing the LCS and CPG branches, aiming to enhance detection robustness and expand representational coverage.
Early fusion techniques commonly relied on straightforward concatenation or performed implicit fusion at the property-graph node embedding stage ~\cite{Linevd,CSGVD},
without explicitly modeling fine-grained semantic correspondences between sequential and structural features, which constrains the expressiveness of the fused representations. 
Inspired by multimodal attention frameworks in fields such as vision-language modeling, more recent approaches incorporate attention-driven fusion modules to facilitate fine-grained cross-modal interactions. In particular, cross-attention mechanisms ~\cite{21CrossFuse} allow sequence-based representations to selectively attend to structural information encoded in the CPG, thereby enhancing alignment and complementarity across modalities. This line of research directly motivates the cross-attention-driven fusion design in our proposed CAFFNet.
It is worth noting that, beyond explicit sequence-graph fusion approaches, recent studies have also explored LLM-based vulnerability detection methods~\cite{12Knighter,39sheng2025llms}. These methods typically rely on prompt-based or zero-shot learning and focus on language-level reasoning, without explicitly constructing or leveraging program-structure representations.

\subsection{Branch Aggregation and Sample-Aware Weighting}
After extracting modality-specific and fused features, an effective aggregation mechanism is required to integrate multi-branch predictions based on sequential, structural, and fused features.
Since the vulnerability patterns present in different datasets may be more discriminatively captured by different modal features, several gating-based and weighting frameworks ~\cite{28,29,22Vul-LMGNNs}
have explored weighted strategies to combine prediction results, aiming to better balance the roles of sequential- and structural-based features at the prediction stage. However, approaches that weight only two modality branches (i.e., sequential and structural branches) make limited use of information from fused representations and thus struggle to fully capture the complementary effects of different semantic branches during decision making. Inspired by prior studies~\cite{29,22Vul-LMGNNs}, we design the SaW strategy, which extends existing dual-branch designs by introducing an additional fused-feature branch to provide complementary cross-modal information. 

\subsection{Research Gap and Our Positioning}
Overall, while sequence-based pre-trained encoders and GNN-based structural models each deliver meaningful advances in code vulnerability detection, neither modality alone is sufficient to capture the multifaceted semantic and structural characteristics of real world source code.
Existing feature fusion methods have limited capability to model deep cross-modal interactions. Moreover, dual-branch aggregation strategies can, to some extent, constrain the effective utilization of fused representations.
To address these issues, we develop FusionVul, which follows a three-stage methodological paradigm encompassing modality-specific feature extraction, cross-attention-based multimodal fusion, and sample-aware weighted aggregation. In this framework, UniXcoder and GGNN extract complementary textual and structural representations, CAFFNet enables explicit cross-modal alignment and information interaction, while SaW integrates predictions across branches, thereby improving detection robustness and discriminative capability.
\section{Proposed Methodology}\label{sec3}
\begin{figure*}[t]
\centering
\includegraphics[width=\textwidth]{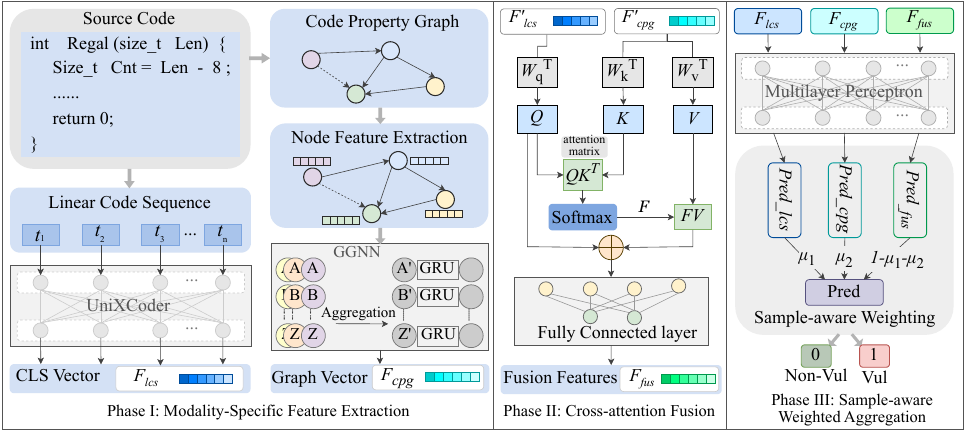}
\caption{Overall framework of FusionVul.}
\label{fig1}
\end{figure*}

FusionVul is designed as an integrated framework that combines sequential and structural perspectives for code semantics modeling. The overall architecture is illustrated in Figure ~\ref{fig1}, while the following subsections elaborate on each component in detail, including input representation, modality-specific feature extraction, cross-attention feature fusion, and sample-aware weighted aggregation.

\subsection{Input Representation}
\subsubsection{Linear Code Sequence}
To capture sequential program syntactic, 
we employ UniXcoder as the sequential-syntactic encoder, as it captures contextual dependencies within linear token sequences and implicitly encodes the syntactic information represented by the AST~\cite{16Unixcoder}.
Each input program fragment is first transformed into a LCS, which serves as the textual modality in our multimodal framework. To maintain efficiency without compromising semantic coverage,
we use the tokenizer provided by UniXcoder to tokenize each function-level code snippet $C$ into a sequence of subword-level tokens, including identifiers, keywords, operators, and literals, and organize these tokens in their linear order in the source code to form the LCS.
Specifically, the tokenizer maintains a fixed vocabulary defined during UniXcoder’s pre-training, and all code tokens are mapped to entries in this vocabulary to ensure compatibility with the model’s embedding layer. For the small number of code snippets whose lengths exceed UniXcoder’s maximum input length, we truncate the sequence to the first 512 tokens. This truncation strategy follows the original UniXcoder evaluation, and all baselines in our experiments are subject to the same length constraint.
Formally, the LCS is obtained via equation (1).
\begin{equation}
    LCS = \mathrm{Tokenizer}(C) = \left\{ t_{1}, t_{2}, \ldots, t_{n} \right\}
\end{equation}

where Tokenizer($\cdot$) denotes a standard lexical analyzer, \textit{C} represents an individual source-code snippet, and the resulting token sequence is directly fed into UniXcoder for embedding and feature extraction. A schematic illustration of the process is provided in Figure~\ref{fig2}(II).

\subsubsection{Code Property Graph}
\begin{figure*}[t]
\centering
\includegraphics[width=\textwidth]{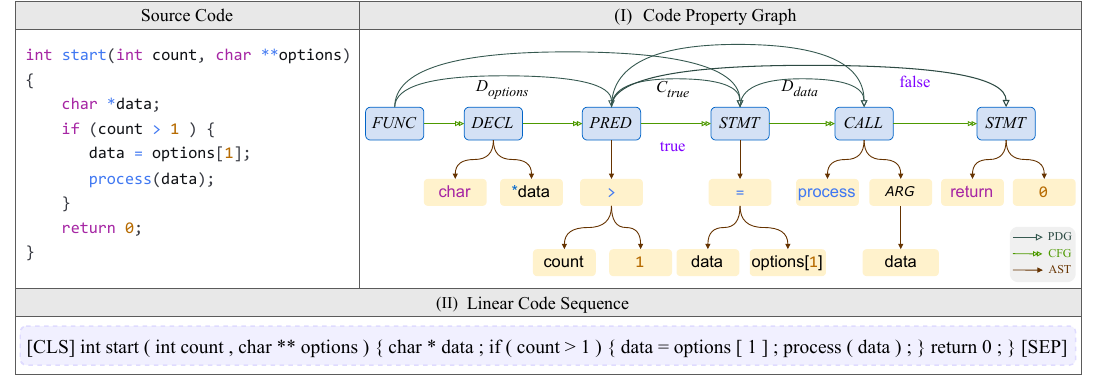}
\caption{LCS and CPG for the example source code.}
\label{fig2}
\end{figure*}

To complement the linear textual representation, we adopt the CPG as a comprehensive structural representation. The CPG integrates three fundamental program-analysis artifacts: the AST, the Control Flow Graph (CFG), and the Program Dependence Graph (PDG). 
This unified formulation allows the model to jointly capture behavioral semantics, including control-flow relations and data-dependence paths, which are essential for identifying vulnerabilities that are related to data flow or driven by control flow. 
In this study, we use Joern, a static analysis and code property extraction framework, to convert each code snippet into its corresponding CPG.
Joern does not impose a fixed upper bound on the length of the input code and is able to analyze the code snippets without truncation.
The generation process includes lexical parsing, AST construction, CFG extraction, and derivation of data-dependency edges, as shown in Figure~\ref{fig2} (I). 

Formally, a CPG is represented as $G = (V, E_T, E_C, E_D)$, where $V$ denotes the set of graph nodes, and $E_{T}$, $E_{C}$ and $E_{D}$ correspond to syntax, control-flow, and data-dependence edges, respectively. For compatibility with the GGNN architecture, we initialize each node $v \in V$ with a feature vector $h_v^{(0)} \in \mathbb{R}^{d_v}$. All node features are stacked to form the initial feature matrix $H^{(0)}$, as defined in equation (2).
\begin{equation}
    H^{(0)} = [h_{1}^{(0)},\, h_{2}^{(0)},\, \ldots,\, h_{|V|}^{(0)}]^{T}
    \in \mathbb{R}^{|V| \times d_{v}}
\end{equation}

where $d_v$ is the node feature dimension, $|V|$ denotes the number of nodes in the CPG. The edge types are represented by an adjacency matrix set $\{A_e\} = \{A_T, A_C, A_D\}$, which together form a tensor $L$ used as input to the GGNN, as expressed in equation (3).
\begin{equation}
    L = \left(H^{(0)},\, \{A_{e}\}_{e \in E}\right)
\end{equation}

Where $E$ denotes the set of edge types, and each $e \in E$ corresponds to a specific relational edge in the CPG. This representation provides a comprehensive foundation for modeling node-level contextual dependencies, which is essential for accurate vulnerability detection.

\subsection{Modality-Specific Feature Extraction}
Following the construction of the LCS and CPG modalities, FusionVul employs UniXcoder and GGNN as complementary encoders to capture multi-level semantic and structural characteristics of program fragments. UniXcoder focuses on learning sequential syntactic representations from tokenized code, whereas GGNN captures relational dependencies and control- and data-flow interactions embedded within the CPG, enabling complementary modeling of syntactic code representations and behavioral semantics. The overall modality-specific feature extraction process is illustrated in Figure ~\ref{fig1} (Phase I).

\subsubsection{Sequence Feature Extraction with UniXcoder}
To encode the lexical and syntactic features of the input LCS, FusionVul utilizes the bidirectional Transformer encoder within UniXcoder.
In our implementation, the pre-trained UniXcoder is used as a fixed feature extractor, with its parameters frozen during training. Given a tokenized sequence $LCS ={\{t_{1},t_{2},\ldots,t_{n}\}}$, each token $t_i$ is mapped to an embedding vector $x_i \in \mathbb{R}^d$ by a learnable embedding layer. These token embeddings are subsequently processed through stacked multi-head self-attention layers, allowing the encoder to model long-range dependencies, control-flow cues, and contextual correlations that emerge throughout the sequence.

After passing through multiple Transformer blocks, the hidden vector associated with the special classification token [CLS] is extracted as the holistic representation of the input code snippet. This vector, denoted as $F_{lcs}$, encapsulates both syntactic patterns and higher-level semantic relationships inferred from global attention interactions. The representation can be formalized as equation (4).
\begin{equation}
    F_{\mathrm{lcs}} = \mathrm{UniXcoder}\left(\{t_{1}, t_{2}, \ldots, t_{n}\}\right)
\end{equation}

Where $F_{lcs} \in \mathbb{R}^{d_s}$ denotes the sequence-level feature embedding produced by the encoder. 

\subsubsection{Structural Feature Extraction with GGNN}
For the structural modality, we adopt the GGNN to model the rich relational information encoded in the CPG. Given the CPG tensor representation $L$, the GGNN performs iterative message passing to update node embeddings. At each iteration $t$, the aggregated message for node $v$ is computed by accumulating transformed features from its relation-specific neighbors. The calculation of aggregated messages is as shown in equation (5).
\begin{equation}
    m_{v}^{(t)} = 
    \sum_{e \in E} \sum_{u \in N_{e}(v)}
        A_{e}[v,u]\, M_{e} \, h_{u}^{(t-1)}
\end{equation}

Here $E$ represents the set of edge types in the CPG, and $u \in N_e(v)$ denotes the neighbors of node $v$ under edge type $e$, and $A_e[v,u]$ is the adjacency weight between nodes $v$ and $u$ for edge type $e$,
i.e., an element of the adjacency matrix $A_e$, which is determined by the CPG. $A_e[v,u]$ determines whether information is propagated from node $u$ to node $v$. Specifically,
If the CPG contains a directed edge of type $e$ from $u$ to $v$, then $A_e[v,u] = 1$, the representation of node $u$ is propagated to node $v$; otherwise, $A_e[v,u] = 0$, the representation of node $u$ does not contribute to the message aggregation of node $v$.
$M_e \in \mathbb{R}^{d\times d}$ is a relation-specific linear transformation matrix, $h_v^{(t)}$ denotes the hidden state of node $v$ after the $t$-th iteration, and the initial states $h_v^{(0)}$ is derived from the node feature vectors in the CPG. 

After message aggregation, node states are updated using a Gated Recurrent Unit (GRU), which integrates the previous state and newly aggregated messages, the node states update process is shown in equation (6).
\begin{equation}
    h_{v}^{(t)} = \mathrm{GRU}\!\left(h_{v}^{(t-1)},\, m_{v}^{(t)}\right)
\end{equation}

Through $T$ propagation iterations, the GGNN gradually integrates multi-hop structural dependencies and propagates information along syntax, control-flow, and data-dependence edges.
Since different nodes often contribute unequally to vulnerability detection, nodes associated with security-critical operations (e.g., API calls and data-flow operations) typically play a more important role.
Accordingly, we apply an attention-based Readout operation over all node embeddings.
The intention behind this mechanism is to encourage the model to assign higher weights to security-critical nodes during graph-level aggregation. We clarify that this represents a design motivation rather than a guarantee of model behavior.
This attention mechanism is implemented as a single-head, sigmoid-gated additive attention scheme without Softmax normalization that weights node contributions, and the final graph representation is obtained as a weighted sum of all node features. Ultimately, the acquisition of CPG features is shown in equation (7).
\begin{equation}
    F_{\mathrm{cpg}} = \mathrm{Readout}\left(\{h_{v}^{T} \mid v \in V\}\right)
\end{equation}

Where Readout($\cdot$) denotes the attention-based aggregation function, $F_{cpg} \in \mathbb{R}^{d_g}$ denotes the final structural representation of the code snippet.

The two feature vectors, $F_{lcs}$ from UniXcoder and $F_{cpg}$ from GGNN, each encode complementary aspects of the input program. These representations form the basis of multimodal feature alignment and fusion in the subsequent CAFFNet. By maintaining independent but complementary extraction pathways, FusionVul preserves modality-specific features that are essential for accurate and robust vulnerability detection.

\subsection{Cross-Attention Feature Fusion}
Following feature extraction, the sequential representation $F_{lcs}$ and the structural representation $F_{cpg}$ serve as two complementary modalities that characterize distinct aspects of the underlying program semantics. To explicitly model their mutual relevance and enhance multimodal coherence, FusionVul incorporates a CAFFNet, which aligns and integrates the two modalities within a unified latent space.

\subsubsection{Feature Dimension Alignment}
Since $F_{lcs} \in \mathbb{R}^{d_s}$ and $F_{cpg} \in \mathbb{R}^{d_g}$ are produced by different encoders, their feature scales and distributions may differ.
To mitigate inconsistencies arising from encoder heterogeneity, CAFFNet first projects the two vectors into a shared feature space with the same dimensionality and a comparable representation scale. Specifically, $F_{lcs}$ and $F_{cpg}$ are independently transformed via linear projection layers with bias terms, mapping them into the shared feature space.
This preliminary alignment ensures that subsequent attention interactions operate on a compatible representation scale. The transformation process is shown in equation (8).
\begin{equation}
\begin{aligned}
    F'_{\mathrm{lcs}} &= W_{s}F_{\mathrm{lcs}} + b_{s}, \\
    F'_{\mathrm{cpg}} &= W_{g}F_{\mathrm{cpg}} + b_{g}
\end{aligned}  
\end{equation}

where $W_s \in \mathbb{R}^{d_f \times d_s}$ and $W_g \in \mathbb{R}^{d_f \times d_g}$ are learnable transformation matrices
that project the feature vectors $F_{lcs}$ and $F_{cpg}$ into a unified feature dimension, and $b_s, b_g \in \mathbb{R}^{d_f}$ are learnable bias terms
that compensate for distributional shifts across modalities, $F'_{lcs}$ and $F'_{cpg}$ are the feature vectors mapped to the unified dimension $d_f$. This mapping aligns the distributions of the LCS and CPG features within a shared latent space, providing a consistent representation for subsequent feature fusion.

\subsubsection{Feature Fusion}
After feature alignment, the resulting features $F'_{lcs}$ and $F'_{cpg}$ may contain both task-relevant semantics for vulnerability detection and task-irrelevant information. To mitigate the interference from irrelevant dimensions, CAFFNet applies a learnable linear transformation to reweight $F'_{lcs}$ and $F'_{cpg}$ enhancing discriminative information while suppressing irrelevant components, thereby projecting the features into a task-relevant representation space.

As illustrated in Figure \ref{fig1} (Phase II), the sequential feature $F'_{lcs}$ is mapped to the Query ($Q$) to represent the current sequential context, while the structural feature $F'_{cpg}$ is mapped to the Key ($K$) and Value ($V$).
Here, $K$ is used to select structural information that is relevant to the sequential features, and $V$ provides the corresponding selected structural representations.
The projection process is shown in equation (9).
\begin{equation}
\begin{aligned}
    Q &= F'_{\mathrm{lcs}} W_{Q}, \\
    K &= F'_{\mathrm{cpg}} W_{K}, \\
    V &= F'_{\mathrm{cpg}} W_{V}
\end{aligned} 
\end{equation}

Where $W_Q$, $W_K$ and $W_V$ are learnable weight matrices
that perform linear representation transformations during feature mapping, projecting the $F'_{lcs}$ and $F'_{cpg}$ into a representation space used for subsequent attention computation. Using Scaled Dot-Product Attention, we compute similarity scores between $Q$ and $K$ to measure how strongly the sequence features attend to the structural features. The attention weights are normalized via the Softmax function, as shown in equation (10).
\begin{equation}
    F = \mathrm{Softmax}\!\left( 
        \frac{Q \cdot K^{T}}{\sqrt{d_{f}}}
    \right)
\end{equation}

Here, $(Q \cdot K^T)$ denotes the dot product between Query and Key, reflecting the relevance between sequence and structural representations. The scaling factor $\sqrt{d_f}$ stabilizes gradients by preventing large magnitude values in high-dimensional spaces. The resulting attention weights $F$ indicate the degree to which each structural feature contributes to the fused representation. Next, the weighted response vector is computed by applying the attention weights to the $V$ matrix, as shown in equation (11).
\begin{equation}
    F' = F \cdot V
\end{equation}

Finally, the resulting response vector $F'$
represents a weighted sum of the selected structural information. It is linear and primarily dominated by structural features. To further integrate sequential semantics with the structural responses, we concatenate $F'$ with $Q$
to form a linear fused representation that combines sequential semantics and structural responses, followed by a fully connected layer that projects the concatenated features into a unified fused feature space.
Since this fusion process mainly consists of stacked linear transformations, we further introduce a non-linear activation function to capture higher-order non-linear interactions between sequential and structural information, thereby enhancing the expressiveness of the fused feature $F_{fus}$.
The process can be expressed as equation (12). 
\begin{equation}
    F_{\mathrm{fus}} = \mathrm{ReLU}\left(
        W_{0}[F' \parallel Q] + b_{0}
    \right)
\end{equation}

Where $W_0$ and $b_0$ denote the learnable parameters of the feed-forward layer, and ReLU($\cdot$) is the activation function. The resulting $F_{fus}$ serves as the final fused representation.

By explicitly establishing cross-modal correspondences, CAFFNet enables sequence features to selectively attend to structural cues. When semantic associations between modalities are strong, for instance when control-flow or data-flow dependencies are reflected in the linearized token sequence, the learned attention weights amplify the relevant structural signals. Conversely, when modality alignment is weak, the attention mechanism suppresses irrelevant structural noise. This fusion strategy yields a semantically richer fused representation that captures both lexical semantics and graph-structured program dependencies, thereby enhancing the effectiveness of subsequent vulnerability prediction.

\subsection{Sample-Aware Weighted Aggregation}
To enhance the robustness of FusionVul across heterogeneous code samples, we incorporate a SaW strategy in the prediction phase, as illustrated in Figure ~\ref{fig1} (Phase III).
SaW is designed as a weighted aggregation scheme based on three prediction branches, namely the sequential, structural, and fused branches. The final prediction is obtained by aggregating the outputs of these branches using predefined weighting coefficients.
This mechanism aims to preserve and combine complementary information from different semantic perspectives, thereby enhancing the thorough utilization of features and improving the generalization capability of the FusionVul.

\subsubsection{Branch-Independent Prediction}
In the first stage, each feature branch independently performs binary classification. Specifically, three Multilayer Perceptrons (MLPs) are applied to the sequence-level representation $F_{lcs}$, the structural representation $F_{cpg}$, and the fused representation $F_{fus}$. Each MLP outputs a probability distribution over the two vulnerability classes, where non-vulnerable and vulnerable are labeled as 0 and 1. This procedure yields three prediction tensors: $Pred_{lcs}, Pred_{cpg}, Pred_{fus} \in \mathbb{R}^{\mathrm{B}\times2}$, where B denotes the batch size (B=8 in our experiments). The first column of each tensor corresponds to the predicted probability of the non-vulnerable class, and the second column corresponds to the predicted probability of the vulnerable class.

\subsubsection{Multi-Branch Weighted Aggregation}
In the second stage, SaW aggregates the predictions from each branch using weighting coefficients. Two coefficients, $\mu1$ and $\mu2$, control the relative importance of the sequence branch $Pred_{lcs}$ and the structural branch $Pred_{cpg}$ respectively. The fused branch $Pred_{fus}$ is assigned the remaining weight $1-(\mu1 + \mu2)$. The final prediction is computed as equation (13).
\begin{equation}
    \mathrm{pred} = 
    \mu_{1} \cdot \mathrm{Pred}_{\mathrm{lcs}} 
    + \mu_{2} \cdot \mathrm{Pred}_{\mathrm{cpg}} 
    + \left(1 - (\mu_{1} + \mu_{2})\right)
        \cdot \mathrm{Pred}_{\mathrm{fus}}
\end{equation}

Where $\mu1, \mu2 \in [0, 1]$ and $(\mu1+\mu2) \leq 1$ represent the weights assigned to the LCS, CPG, and fused branches, respectively, ensuring that the total contribution sums to one. The final output $Pred \in \mathbb{R}^{\mathrm{B}\times2}$ denotes the weighted probability distribution over the two prediction classes.

\subsubsection{Discussion on Parameter Settings}
Different values of $\mu1$ and $\mu2$ reflect different emphases on the semantic perspectives. For example:
\begin{itemize}
\item when $\mu1=1$ and $\mu2=0$, the model relies solely on the sequence-level branch.
\item when $\mu1=0$ and $\mu2=1$, prediction depends exclusively on structural information.
\end{itemize}
In more general scenarios, neither coefficient is fixed at extreme values. 
Intermediate values allow the aggregation scheme to combine modality-specific representation with the fused representation, where the fusion branch provides complementary information. From a design perspective, increasing $\mu2$ while reducing $\mu1$ allows greater emphasis on structural cues. Conversely, when the LCS exhibits rich semantic content, a larger $\mu1$ is more appropriate.

The SaW strategy maintains the sequential-based, structural-based, and fusion-based representations as independent prediction branches and performs weighted aggregation over the multi-branch predictions. This design allows the final decision to integrate complementary information from multiple semantic perspectives, thereby improving FusionVul’s detection robustness across diverse vulnerability patterns.

\section{Evaluation}\label{sec4}
In this section, we describe the experimental setup and the four datasets used for evaluation. To comprehensively assess the effectiveness of the proposed FusionVul framework, we formulate four research questions (RQs) and address them through comparative experiments involving six state-of-the-art vulnerability detection approaches and four widely adopted pre-trained code models.

\textbf{RQ1}: Compared with a single-model approach, what advantages does our hybrid model, which combines a pre-trained model with a GGNN, offer for vulnerability detection?

\textbf{RQ2}: How effective are the LCS feature $F_{lcs}$ and the CPG structural feature $F_{cpg}$ in predicting vulnerabilities within the detection task?

\textbf{RQ3}: How do variations in the weighting coefficients $\mu1$ and $\mu2$ within the SaW strategy affect detection performance?

\textbf{RQ4}: How does the performance of the pre-trained model UniXcoder compare with that of other popular pre-trained models when applied to vulnerability detection?

These research questions collectively aim to evaluate the design rationality and component effectiveness of FusionVul.

All experiments were conducted on a server equipped with an NVIDIA RTX 4090 GPU (24 GB VRAM) and 90 GB system memory. The software environment includes NVIDIA driver 550.78 with CUDA 12.4, Python 3.10.12, and PyTorch 2.4.1. We consistently used Joern v1.1.1700 for CPG generation.

\subsection{Dataset Preparation}
To enhance the realism and applicability of our evaluation,
we select datasets according to the following criteria. First, the datasets should be constructed from real-world vulnerable code, rather than synthetic or rule-generated samples, so as to reflect practical security analysis scenarios. 
Second, the datasets should provide function-level vulnerability annotations, where each sample corresponds to a single function labeled as vulnerable or non-vulnerable.  
Finally, we prioritize datasets that are widely adopted in prior studies, allowing fair comparison with existing approaches while covering diverse vulnerability patterns. 
Based on these criteria, we select the following four datasets to evaluate our method.

Devign ~\cite{17Devign} is constructed from vulnerability-fixing commits of four open-source projects. By contrasting code changes before and after the fixes, the dataset extracts samples at the function level and labels them as vulnerable or non-vulnerable functions.

ReVeal ~\cite{26DeepReveal} mainly comprises function-level samples derived from the vulnerability reporting and fixing processes of real-world projects such as Chromium and the Debian Kernel. The samples are curated and labeled based on vulnerability-related code evolution information.

SVulD ~\cite{36} is a challenging vulnerability detection dataset derived from a filtered version of Big-Vul, which is built based on CVE records and real-world open-source projects, with samples formed at the function level. Its positive and negative samples often share highly similar surface syntax, while differing in underlying semantic behaviors.

DiverseVul ~\cite{1Diversevul} is a recently proposed large-scale vulnerability detection dataset. It is also organized at the function level. This dataset is large in scale and covers a wide range of vulnerabilities, making it suitable for evaluating the model’s robustness and generalization capability in cross-project and complex-vulnerability scenarios.

All four benchmarks treat a function-level code snippet as one sample, labeled as either vulnerable or non-vulnerable. Our experimental setting follows the same evaluation granularity adopted by most prior studies based on these datasets.
Table~\ref{tab1} provides the statistical characteristics of these datasets.
Where \textit{Total} denotes the total number of functions, \textit{Vul} and \textit{Non-Vul} denote the numbers of vulnerable and non-vulnerable functions, respectively, and \textit{Ratio} represents the proportion of vulnerable functions.
In addition to sample counts, Table~\ref{tab1} reports two properties that reflect dataset diversity: the number of source projects (\textit{Projects}) and the number of distinct CWE categories covered (\textit{CWEs}). The \textit{Projects} column indicates the breadth of code origins, and the \textit{CWEs} column quantifies the variety of vulnerability types represented in each dataset. 
As shown in the table, SVulD and DiverseVul are sourced from 348 and 797 projects respectively, indicating substantially greater source diversity than Devign and Reveal, both of which are derived from only 2 projects.
Furthermore, SVulD and DiverseVul provide explicit CWE annotations covering 91 and 150 vulnerability categories respectively, a characteristic that reflects the design emphasis of these two benchmarks on broad vulnerability-type coverage, as documented in their original papers~\cite{36,1Diversevul}. 

\begin{table*}
\centering
\caption{The statistical characteristics of the datasets.\label{tab1}}
\begin{threeparttable}
\begin{tabular*}{\linewidth}{@{\extracolsep{\fill}}lccrrrr@{}}
\toprule  
\textbf{Dataset} 
& \multicolumn{1}{c}{\textbf{Projects}}  
& \multicolumn{1}{c}{\textbf{CWEs}}  
& \multicolumn{1}{c}{\textbf{Total}}  
& \multicolumn{1}{c}{\textbf{Vul}}  
& \multicolumn{1}{c}{\textbf{Non-Vul}}  
& \multicolumn{1}{c}{\textbf{Ratio} }  \\
\midrule
Devign  & 2 & N/A\tnote{*} &26,037 &11,888 &14,149 &$45.66\%$ \\
ReVeal  & 2 & N/A\tnote{*} &18,169 &1,664 &16,505 &$9.16\%$ \\  
SVulD   & 348 & 91 &29,256 &5,260 &23,996 &$17.97\%$ \\
DiverseVul   & 797 & 150 & 349,437 &  18,945 & 330,492 & $5.42\%$ \\
\bottomrule
\end{tabular*}
\begin{tablenotes}
    \footnotesize
    \item[*] N/A indicates that the original dataset does not provide CWE annotations, precluding a reliable count of vulnerability types.
\end{tablenotes}
\end{threeparttable}
\end{table*}

The four datasets collectively provide a comprehensive evaluation environment for defect detection. We then randomly split each deduplicated dataset into training, validation, and test sets with a 6:2:2 ratio. We do not apply class-balancing techniques. Instead, we follow the original settings of the adopted datasets and the common practice in prior studies using the same datasets by preserving the original class distribution. All experiments in this study are conducted under the same data splits and class distributions.

\subsection{Dataset Characterization}
Given the potential influence of dataset characteristics on vulnerability detection performance, we systematically define and quantify dataset properties from two complementary perspectives: (i) the structural complexity of the CPGs, and (ii) the separability in the embedding space between vulnerable and non-vulnerable samples.
To quantify the complexity of the structure and the distinguishability between positive and negative samples, we defined four CPG levels and two embedding levels of metrics for each dataset. The definitions of all metrics are shown in Table~\ref{tab2}.

\begin{table*}
\centering
\caption{Definitions of metrics.\label{tab2}}
\begin{threeparttable}
\begin{tabular*}{\linewidth}{@{\extracolsep{\fill}} m{0.2\linewidth} m{0.8\linewidth} @{}}
\toprule  
\textbf{Metric}  
& \multicolumn{1}{l}{\textbf{Definition}}  
\\
\midrule
$|V|$   & Count of all CPG nodes.  \\
$|E|$  & Count of all CPG edges.  \\  
$|E|/|V|$ & Edge-to-Node ratio.  \\
Max.depth   & Longest path from root in the AST sub-graph. \\
FDR & Ratio of inter-class distance to intra-class variance between positive and negative samples in the embedding space. \\
MMD & Distance between the distributions of positive and negative samples in the embedding space. \\
\bottomrule
\end{tabular*}
\end{threeparttable}
\end{table*}

\begin{figure*}[t]
\centering
\includegraphics[width=\textwidth]{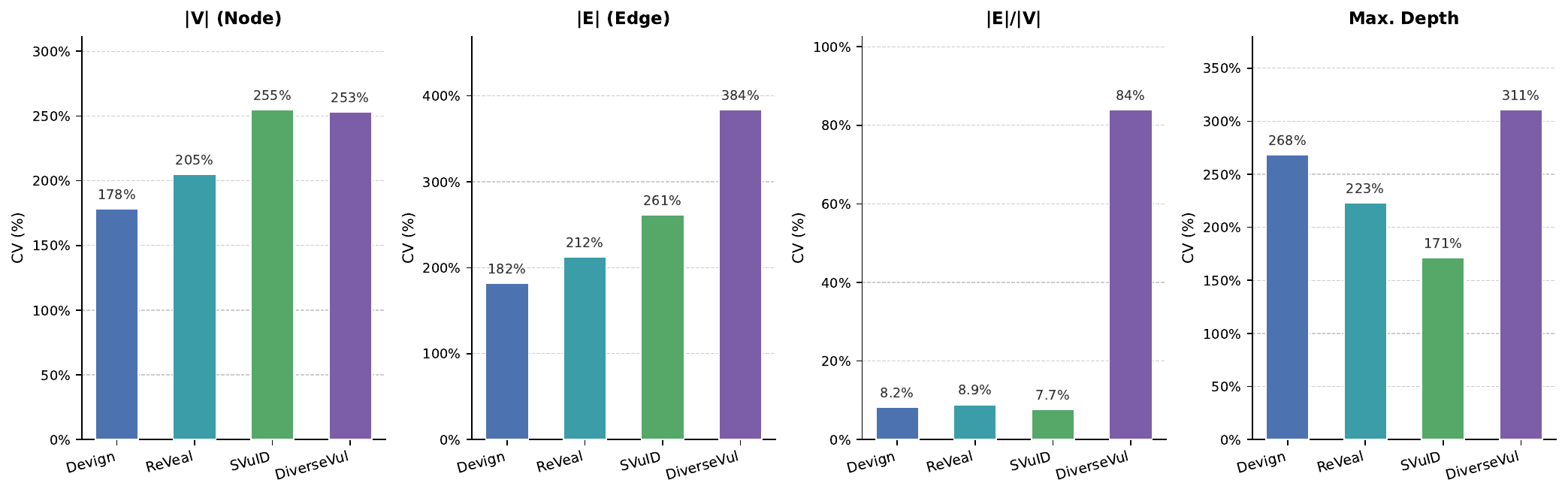}
\caption{Coefficient of variation (CV) of CPG structural metrics.}
\label{fig3}
\end{figure*}

From the structural perspective, $|V|$ and $|E|$ measure the scale of the CPG, reflecting the size of the program representation in terms of nodes and edges, respectively. Larger values typically indicate richer semantic information.
The edge-to-node ratio ($|E|/|V|$)  captures the connection density of the CPG, indicating the average number of relationships associated with each node. A higher ratio suggests denser interactions among program elements, implying increased structural complexity.
Max.depth reflects the syntactic nesting depth of the program. A larger depth indicates more deeply nested structures, often corresponding to higher logical complexity and increased difficulty for analysis. 
For each dataset, we report the mean ($m$), standard deviation ($\sigma$), and coefficient of variation (CV) of each CPG-level metric. Table~\ref{tab3} presents the CPG structural statistics.
Figure~\ref{fig3} shows the differences in the coefficients of variation among the four metrics.

\begin{table*}[htbp]
\scriptsize
\centering
\caption{CPG structural statistics of four datasets.\label{tab3}}
\begin{threeparttable}
\begin{tabularx}{\linewidth}{l *{13}{Y}}
\toprule  
&
\multicolumn{3}{c}{\textbf{$|V|$}} 
& 
\multicolumn{3}{c}{\textbf{$|E|$}} 
&
\multicolumn{3}{c}{\textbf{$|E|/|V|$}} 
& 
\multicolumn{3}{c}{\textbf{Max.depth}} 
\\
\cmidrule(l){2-4}\cmidrule(l){5-7}\cmidrule(l){8-10}\cmidrule(l){11-13}
\textbf{Dataset}
& $m$ & $\sigma$ & CV\tnote{*} & $m$ & $\sigma$ & CV\tnote{*} 
& $m$ & $\sigma$ & CV\tnote{*} & $m$ & $\sigma$ & CV\tnote{*} \\
\midrule
Devign & 196 & 349 & \textbf{178\%}  
& 375 & 681 & \textbf{182\%} 
& 1.82 & 0.15 & \textbf{8.2\%} 
& 10.75 & 28.82 & \textbf{268\%} \\
ReVeal  & 142 & 291 &  \textbf{205\%} 
& 269 & 569 & \textbf{212\%} 
& 1.79 & 0.16 & \textbf{8.9\%} 
& 9.12 & 20.33 & \textbf{223\%} \\  
SVuID  & 149 & 380 & \textbf{255\% }
& 283 & 739 & \textbf{261\%} 
& 1.81 & 0.14 & \textbf{7.7\%} 
& 8.29 & 14.2 & \textbf{171\%} \\
DiverseVul & 373 & 947 & \textbf{253\% }
& 1479 & 5676 & \textbf{384\% }
& 3.96 & 3.31 & \textbf{84\%} 
& 12.5 & 38.9 & \textbf{311\%} \\
\bottomrule
\end{tabularx}
\begin{tablenotes}
    \footnotesize
    \item[*] $ \text{CV}= \sigma/m \times 100\%$. A higher CV indicates greater data dispersion and larger structural variation among functions. 
\end{tablenotes}
\end{threeparttable}
\end{table*}

Figure~\ref{fig3} compares the CV of four CPG structural metrics across datasets. For $|V|$ and $|E|$, Devign and ReVeal show lower CVs (178\%–212\%), while SVulD and DiverseVul are substantially higher (253\%–384\%), indicating that the function size distribution of the former two is more concentrated, whereas that of the latter two is more dispersed. This disparity reflects the broader source diversity of SVulD and DiverseVul, which are derived from 348 and 797 projects respectively, in contrast to the 2-project origins of Devign and ReVeal.  
Regarding $|E|/|V|$, the metric remains stable across Devign, ReVeal, and SVulD (7.7\%–8.9\%), but rises sharply in DiverseVul (84\%), indicating highly variable structural density driven by its wide CWE coverage (150 CWEs). Max.depth exhibits uniformly high CVs (171\%–311\%) across all datasets, reflecting considerable variation in syntactic nesting depth across functions universally. 
Taken together, Devign and ReVeal exhibit relatively stable CPG structural characteristics. SVulD shows mixed behavior, being more dispersed in function size but stable in connection density. In contrast, DiverseVul is the most structurally heterogeneous overall.

From the embedding perspective, \textit{Fisher Discriminant Ratio}(FDR) and \textit{Maximum Mean Discrepancy}(MMD) are used to quantify the separability between vulnerable and non-vulnerable samples. 
Specifically, FDR is a classical criterion derived from Fisher discriminant analysis~\cite{mika1999fisher}, widely used to measure class separability based on statistical characteristics (i.e., mean and variance)~\cite{nips2005fdr}. 
In contrast, MMD is a two-sample test that measures the discrepancy between distributions~\cite{gretton2012MMD}, providing a more general assessment of separability at the distribution level~\cite{gao2015sub}.
Higher values of these two metrics indicate higher separability between positive and negative samples in the embedding space.
For each dataset, we compute these metrics on both LCS embeddings and CPG embeddings, enabling a comparative analysis of separability in terms of sequential and structural features.
The specific results are shown in Table ~\ref{tab4}.

\begin{table}[htbp]
\scriptsize
\centering
\caption{Separability of positive and negative samples across embedding spaces.\label{tab4}}
\begin{tabularx}{\linewidth}{l *{5}{Y}}
\toprule  
&
\multicolumn{2}{c}{\textbf{LCS}} 
& 
\multicolumn{2}{c}{\textbf{CPG}} 
\\
\cmidrule(l){2-3}\cmidrule(l){4-5}
\textbf{Dataset}
& FDR & MMD & FDR & MMD  \\
\midrule
Devign & 21.4 & 15.7 & 6.3  & 4.6  \\
ReVeal  & 24.5 & 19.3 &  8.2 & 9.4 \\  
SVulD  &  4.5 & 8.6 & 27.5 & 18.9  \\
DiverseVul & 7.3 & 9.3 & 16.8 & 12.4 \\
\bottomrule
\end{tabularx}
\end{table}

\begin{figure*}[h]
\centering
\includegraphics[width=\textwidth]{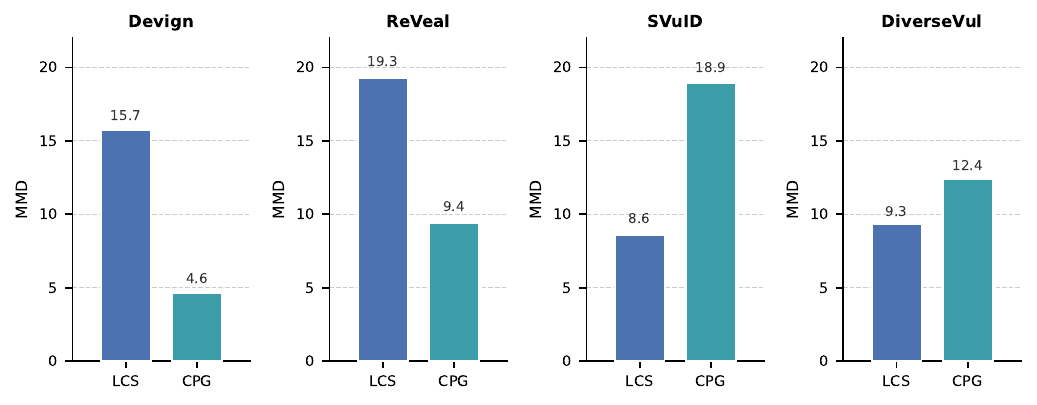}
\caption{Separability comparison of LCS and CPG embeddings on each dataset (MMD).}
\label{fig4}
\end{figure*}

As shown in Table~\ref{tab4}, LCS embeddings exhibit substantially higher separability than CPG on Devign and ReVeal (FDR: 21.4 vs. 6.3 and 24.5 vs. 8.2; MMD: 15.7 vs. 4.6 and 19.3 vs. 9.4), indicating that vulnerability patterns in these datasets are better captured by token-level sequential features than structural dependencies.
In contrast, CPG embeddings dominate on SVulD (FDR: 27.5 vs. 4.5; MMD: 18.9 vs. 8.6), consistent with its characteristic that vulnerable and non-vulnerable samples share similar surface syntax but differ in underlying semantics, making control- and data-flow dependencies more discriminative.
For DiverseVul, neither modality shows a clear advantage (FDR: 7.3 vs. 16.8; MMD: 9.3 vs. 12.4), with overall low separability. 
This reflects the complexity and diversity of DiverseVul, which covers a wide range of vulnerability types (150 CWEs) with heterogeneous structural and semantic characteristics.
Figure~\ref{fig4} further illustrates, in bar chart form, the differences in separability between LCS and CPG embeddings within each dataset.

\subsection{Baseline Comparison (RQ1)}
To ensure a comprehensive and objective evaluation of FusionVul’s performance, the baseline methods were selected based on three principles:
maintaining consistency in function-level granularity, ensuring methodological diversity (encompassing sequence-based, structure-based, and multimodal approaches), and prioritizing approaches with wide adoption in recent literature. Accordingly, six representative methods were chosen as baselines.

LineVul ~\cite{34}: A Transformer-based sequence model that focuses solely on linear code representations and does not incorporate structural graph information.

Devign ~\cite{17Devign}: A GGNN-based method that embeds the LCS implicitly into a CPG, relying exclusively on the graph network for feature propagation.

Reveal ~\cite{26DeepReveal}: Also built upon the GGNN architecture, this method uses the CPG as input and focuses solely on capturing structural dependencies within the code graph.

Vul-LMGNNs ~\cite{22Vul-LMGNNs}: A multimodal approach that employs both a Transformer-based sequence encoder and a graph-based encoder. The framework employs a dual-branch architecture that incorporates both sequential and structural features; however, the fused representations are not aggregated via weighted summation.

SCALE ~\cite{25Scale}: A Transformer-based sequence model that encodes abstract syntax trees and code comments as linearized sequences but does not utilize property graph information.

CSGVD ~\cite{CSGVD}: A multimodal approach that integrates a pre-trained sequence encoder with  a Graph-based encoder, ultimately making predictions using a fused global representation, without employing multi-branch weighted aggregation.

We evaluate all methods on the four datasets. 
For FusionVul, we trained the model for 100 epochs with a batch size of 32, using the Adam optimizer with an initial learning rate of 1e-4, and optimized it with the cross-entropy loss. For all baselines, we followed the hyperparameter settings reported in their original papers or official implementations. When specific parameters were not explicitly provided, we used the default settings in the released code. All methods were re-implemented and evaluated under the same experimental environment.
The comparative results are summarized in Tables~\ref{tab5} and~\ref{tab6}.

\begin{table}[htbp]
\scriptsize
\centering
\caption{Performance comparison on Devign and Reveal datasets.\label{tab5}}
\begin{tabularx}{\linewidth}{l *{12}{Y}}
\toprule  
&
\multicolumn{3}{c}{\textbf{Devign dataset}} 
& 
\multicolumn{3}{c}{\textbf{Reveal dataset}} 
\\
\cmidrule(l){2-4}\cmidrule(l){5-7}

\textbf{Method} & P & R & F1 & P & R & F1 \\
\midrule
LineVul & 61.55 & 48.21 & 54.07  & 48.60 & 56.15 & 49.10 \\
Devign  & 52.90 & 56.37 &  54.58 & 36.65 & 31.55 &  33.91 \\  
Reveal  &  59.80 & 53.71 & 56.59 & 29.90 & 40.91 & 33.87 \\
Vul-LMGNNs & 63.64 & 53.85 & 58.33 & 44.38 & 33.65 & 38.27 \\
SCALE  & 61.88 & \textbf{68.69} &  \textbf{65.11} & \textbf{52.32} & \textbf{78.69} &  \textbf{62.85} \\  

CSGVD  & 62.19 & 58.06 & 60.05 & 46.63 & 41.08 & 43.68 \\  

\textbf{FusionVul} &  \textbf{69.65} & 50.31 & 58.42 & 50.82 & 43.87 & 47.09 \\
\bottomrule
\end{tabularx}
\end{table}

\begin{table}[htbp]
\scriptsize
\centering
\caption{Performance comparison on SVulD and DiverseVul datasets.\label{tab6}}
\begin{tabularx}{\linewidth}{l *{12}{Y}}
\toprule  
&
\multicolumn{3}{c}{\textbf{SVulD dataset}} 
& 
\multicolumn{3}{c}{\textbf{DiverseVul dataset}} 
\\
\cmidrule(l){2-4}\cmidrule(l){5-7}
\textbf{Method} & P & R & F1 & P & R & F1 \\
\midrule
LineVul & 15.95 & \textbf{64.45} & 25.58 
& 11.48 & 10.63 & 11.04\\
Devign  &  9.72 & 50.31 &  16.29 
& 9.35 & 9.22 & 9.28 \\  
Reveal  &  12.92 & 40.08 & 19.31 
& 9.35 & 18.46 & 12.42\\
Vul-LMGNNs  & 57.83 & 45.39 & 50.86 
& 32.08 & 18.56 & 23.52\\
SCALE  &  22.56 & 57.03 &  32.33 
& 14.36 & 9.85 & 11.68\\  
CSGVD  & 55.35 & 43.86 & 48.94 
& 26.45 & 16.94 & 20.65 \\ 
\textbf{FusionVul}  &  \textbf{65.98} & 45.50 & \textbf{53.86} 
& \textbf{32.10} & \textbf{20.64} & \textbf{25.12}\\
\bottomrule

\end{tabularx}
\end{table}

The results in the table show that SCALE achieves higher F1 and Recall than FusionVul on Devign and Reveal, while FusionVul attains the best F1 on SVulD and DiverseVul. Concretely, FusionVul achieves F1 scores of 58.42\% (Devign), 47.09\% (Reveal), 53.86\% (SVulD), and 25.12\% (DiverseVul), ranking first on the latter two datasets.

This result can be explained by the dataset properties quantified in Section 4.2. 
Specifically, as shown in Table~\ref{tab4}, Devign and ReVeal exhibit substantially higher separability in LCS embeddings than in CPG embeddings (MMD: 15.7 vs. 4.6 for Devign; 19.3 vs. 9.4 for ReVeal), indicating that vulnerability patterns in these datasets are more discriminatively captured by token-level sequential features.
In such cases, methods focusing primarily on sequential features, such as SCALE, may exhibit stronger inductive bias, which explains their higher recall and F1 on these two datasets. 
However, SCALE remains a sequence-based, single-modality method. Its performance drops noticeably on SVulD and DiverseVul, where, as shown in Table~\ref{tab4}, 
LCS embeddings exhibit substantially lower separability (LCS MMD: 8.6 and 9.3 respectively), 
while CPG embeddings become more discriminative (CPG MMD: 18.9 and 12.4 respectively).
This suggests that, in these datasets, syntactic differences at the token sequence level are less discriminative, and vulnerability pattern recognition relies more on structural-semantic features.
Furthermore, Table~\ref{tab1} shows the broader coverage of vulnerability types in these two datasets (SVulD: 91 CWEs; DiverseVul: 150 CWEs), reflecting the diversity of vulnerability categories they contain.

In contrast, FusionVul leverages cross-modal feature fusion and a multi-branch aggregation strategy, achieving higher and more stable F1 scores on datasets with greater structural heterogeneity and broader vulnerability-type coverage, such as SVulD and DiverseVul. 
Although its recall on SVulD is lower than that of some baselines, it attains higher precision. 
This precision advantage suggests that cross-modal fusion and multi-branch aggregation help suppress false positives when function size is highly dispersed across the dataset (as reflected by the CV of $|E|$ in Table 3) and when the range of vulnerability types covered is broad (as indicated by the CWE counts reported in Table 1).
The improved F1 score further indicates that the gain in precision compensates for the decrease in recall, reflecting a stronger ability to capture complex and diverse vulnerabilities.

In summary, FusionVul does not achieve the highest F1 across all datasets. On Devign and ReVeal, where LCS embeddings exhibit high separability, SCALE, as a sequence-based method, attains higher recall and F1. In contrast, FusionVul achieves the best F1 on SVulD and DiverseVul, where function size is highly dispersed and the range of vulnerability types covered is broad.
The higher F1 scores on these datasets indicate that FusionVul is suited to complex vulnerability scenarios where structural and semantic diversity are prominent.

\subsection{Comparison of Different Modal Features (RQ2)}
FusionVul employs a multimodal feature fusion strategy for code defect detection. To evaluate the predictive effectiveness of each feature branch, we designed the following two variant models:
\begin{itemize}
\item $\mathrm{FusionVul_{w/oG}}$ removes the GGNN branch and relies solely on the sequence feature $F_{lcs}$ extracted by UniXcoder for prediction.
\item $\mathrm{FusionVul_{w/oU}}$ removes the UniXcoder branch and uses only the graph-structured feature $F_{cpg}$ extracted by GGNN for prediction.
\end{itemize}
We compare the original FusionVul with these two ablated variants across four datasets. The experimental results are summarized in Tables~\ref{tab7} and~\ref{tab8}.

\begin{table}[!h]%
\scriptsize
\centering
\caption{Comparison of the variant methods on Devign and Reveal datasets.\label{tab7}}
\begin{tabularx}{\linewidth}{l *{12}{Y}}
\toprule  
&
\multicolumn{3}{c}{\textbf{Devign dataset}} 
& 
\multicolumn{3}{c}{\textbf{Reveal dataset}} 
\\
\cmidrule(l){2-4}\cmidrule(l){5-7}
\textbf{Method}   
& \multicolumn{1}{c}{\textbf{P}}  & \multicolumn{1}{c}{\textbf{R}}  & \multicolumn{1}{c}{\textbf{F1}} 
& \multicolumn{1}{c}{\textbf{P}}  & \multicolumn{1}{c}{\textbf{R}}  & \multicolumn{1}{c}{\textbf{F1}} \\
\midrule
$\mathrm{FusionVul_{w/oG}}$  & 62.67 & \textbf{52.88} & 57.36 
& \textbf{52.67} & 37.26 & 43.65 \\
$\mathrm{FusionVul_{w/oU}}$  & 46.18 & 11.08 &  17.87  
& 9.06 & 27.23 &  13.60 \\   
\textbf{FusionVul}  &  \textbf{69.65} & 50.31 & 58.42 
& 50.82 & \textbf{43.87} & \textbf{47.09} \\
\bottomrule
\end{tabularx}
\end{table}

\begin{table}[!h]%
\scriptsize
\centering
\caption{Comparison of the variant methods on SVulD and DiverseVul datasets. \label{tab8}}
\begin{tabularx}{\linewidth}{l *{12}{Y}}
\toprule  
&
\multicolumn{3}{c}{\textbf{SVulD dataset}} 
& 
\multicolumn{3}{c}{\textbf{DiverseVul dataset}} 
\\
\cmidrule(l){2-4}\cmidrule(l){5-7}
\textbf{Method}  
& \multicolumn{1}{c}{\textbf{P}}  & \multicolumn{1}{c}{\textbf{R}}  & \multicolumn{1}{c}{\textbf{F1}} 
& \multicolumn{1}{c}{\textbf{P}}  & \multicolumn{1}{c}{\textbf{R}}  & \multicolumn{1}{c}{\textbf{F1}}  \\
\midrule
$\mathrm{FusionVul_{w/oG}}$  & 64.16 & 42.32 & 51.00 
& 30.62 & 18.64 & 23.17 \\
$\mathrm{FusionVul_{w/oU}}$  &  58.04 & 19.62 &  29.33 
& 24.35 & 9.83 & 14.00\\   
\textbf{FusionVul} &  \textbf{65.98} & \textbf{45.50} & \textbf{53.86} 
& \textbf{32.10} & \textbf{20.64} & \textbf{25.12} \\
\bottomrule
\end{tabularx}
\end{table}

To address RQ2, we focus on analyzing the independent predictive capability and complementarity of the LCS feature $F_{lcs}$ and the CPG structural feature $F_{cpg}$ in vulnerability detection. The results in the table show that the variant $\mathrm{FusionVul_{w/oG}}$ achieves an F1-score of 57.36\% on the Devign dataset, while $\mathrm{FusionVul_{w/oU}}$ drops sharply to 17.87\%. This result indicates that the sequential representations captured by the pre-trained model encode more comprehensive global semantics of the code. However, even though $\mathrm{FusionVul_{w/oG}}$ performs relatively well, its F1-score remains below that of the full FusionVul model (58.42\%), demonstrating that relying solely on sequential features is insufficient to represent the structural dependencies essential for accurate vulnerability identification. Furthermore, $\mathrm{FusionVul_{w/oU}}$ consistently shows the weakest performance on the Reveal, SVulD, and DiverseVul datasets. Although $\mathrm{FusionVul_{w/oG}}$ remains stable, its F1-score is 3.44 percentage points lower than FusionVul on Reveal, 2.86 percentage points lower on SVulD, and 2.66 percentage points lower on DiverseVul. This observation indicates that using either LCS or CPG features alone provides limited predictive capability, with LCS-based models achieving stronger baseline performance than CPG-based ones, while their combination leads to a substantial improvement in overall F1.

Across all four datasets, the full FusionVul model consistently achieves the highest F1-scores, further demonstrating the effectiveness and robustness of the synergistic fusion of $F_{lcs}$ and $F_{cpg}$ representations.

\subsection{Influence of SaW Strategy Coefficients (RQ3)}
To investigate the impact of $\mu1$ and $\mu2$ on detection performance, we conduct a series of parameter sensitivity analyses. 
This experiment aims to analyze the performance trend of the proposed method under different weight configurations. To ensure experimental control and interpretability, we fix the dataset and all other influencing factors so that any performance variation can be attributed solely to changes in the weighting parameters. This controlled setting enables a clearer interpretation of the impact of weight coefficients within the SaW mechanism.

We conduct this analysis on the Devign and SVulD datasets.
As characterized in Section 4.2, Devign features a relatively balanced class distribution and a moderate scale, exhibiting token-level sequential features with high discriminability.
In contrast, SVulD represents a more challenging scenario with subtler semantic distinctions and greater structural complexity. 
In our experiments, both $\mu1$ and $\mu2$ are assigned values from the set [0, 0.1, 0.2, 0.3, 0.4, 0.5, 0.6, 0.7, 0.8, 0.9, 1.0], yielding 11 discrete points for each parameter. All parameter combinations are applied according to equation (13), and the results are visualized as a heatmap, as shown in Figure~\ref{fig5}.

\afterpage{%
\begin{figure*}[p]
\centering
\vspace{-0.3cm}
\begin{subfigure}{.5\textwidth}
    \centering
    \includegraphics[width=\linewidth,height=0.33\textheight,keepaspectratio]{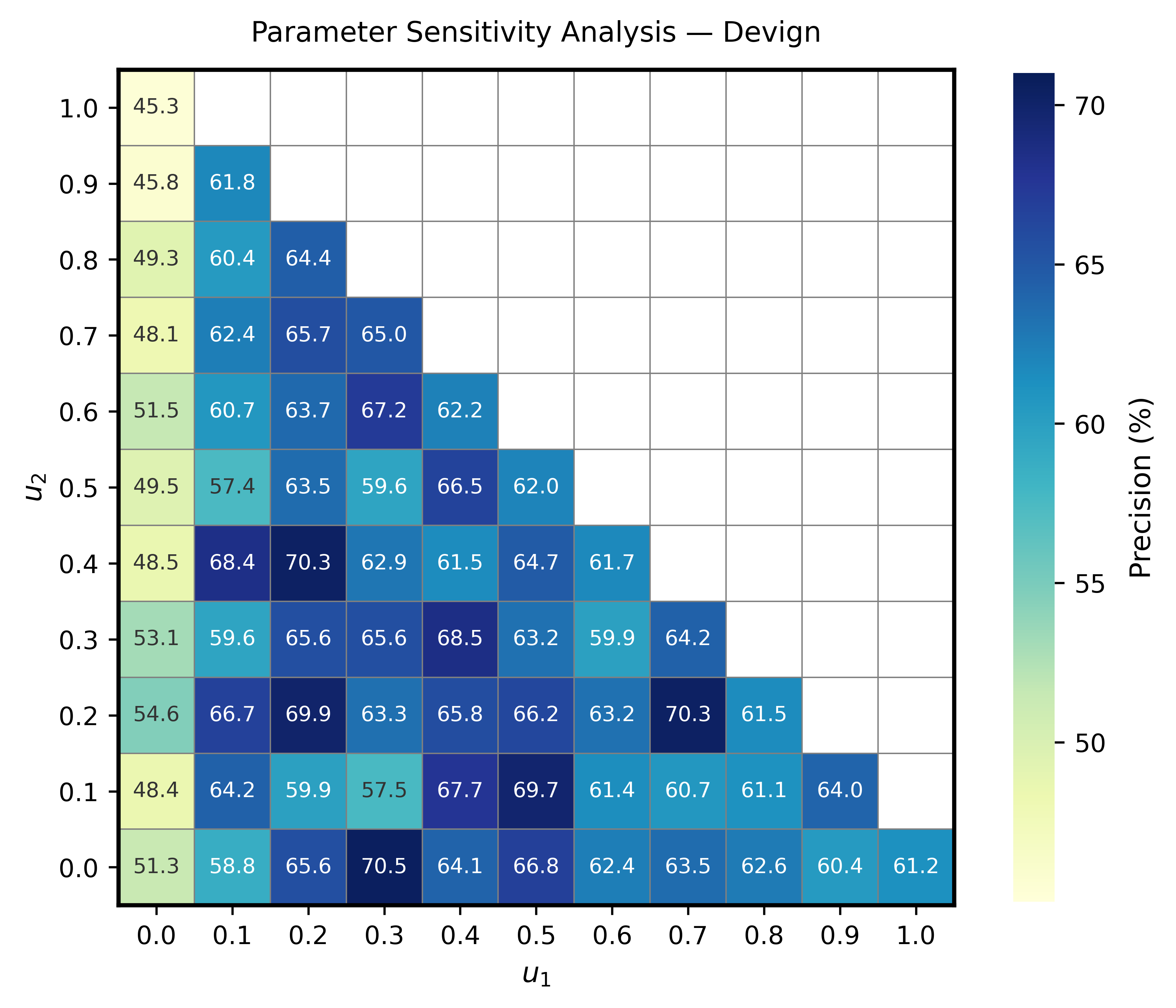}
    \vspace{-0.4cm}
    \label{fig:devign_p}
\end{subfigure}
\hspace{-0.02\textwidth}
\begin{subfigure}{.5\textwidth}
    \centering
    \includegraphics[width=\linewidth,height=0.33\textheight,keepaspectratio]{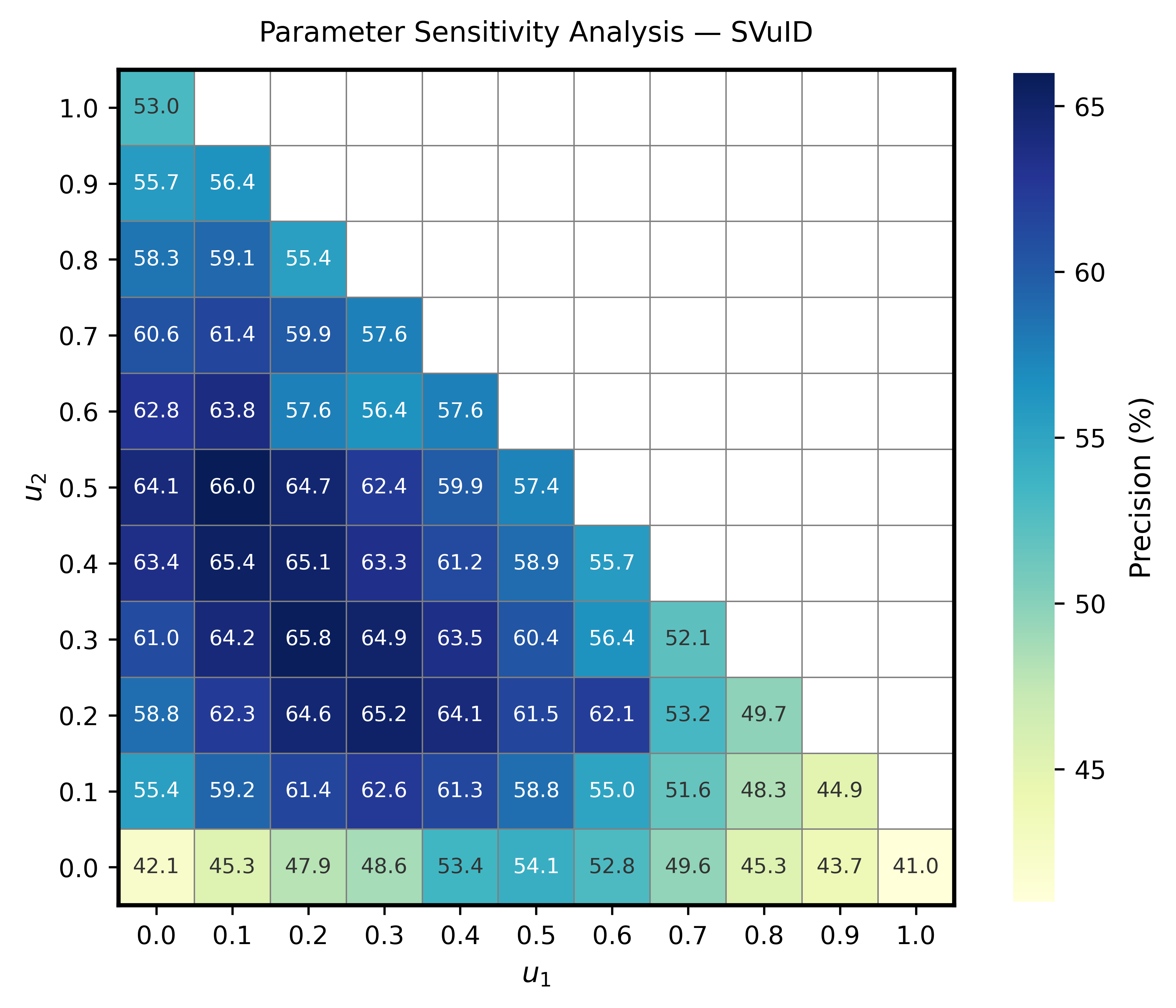}
    \vspace{-0.4cm}
    \label{fig:svuid_p}
\end{subfigure}

\begin{subfigure}{.5\textwidth}
    \centering
    \includegraphics[width=\linewidth,height=0.33\textheight,keepaspectratio]{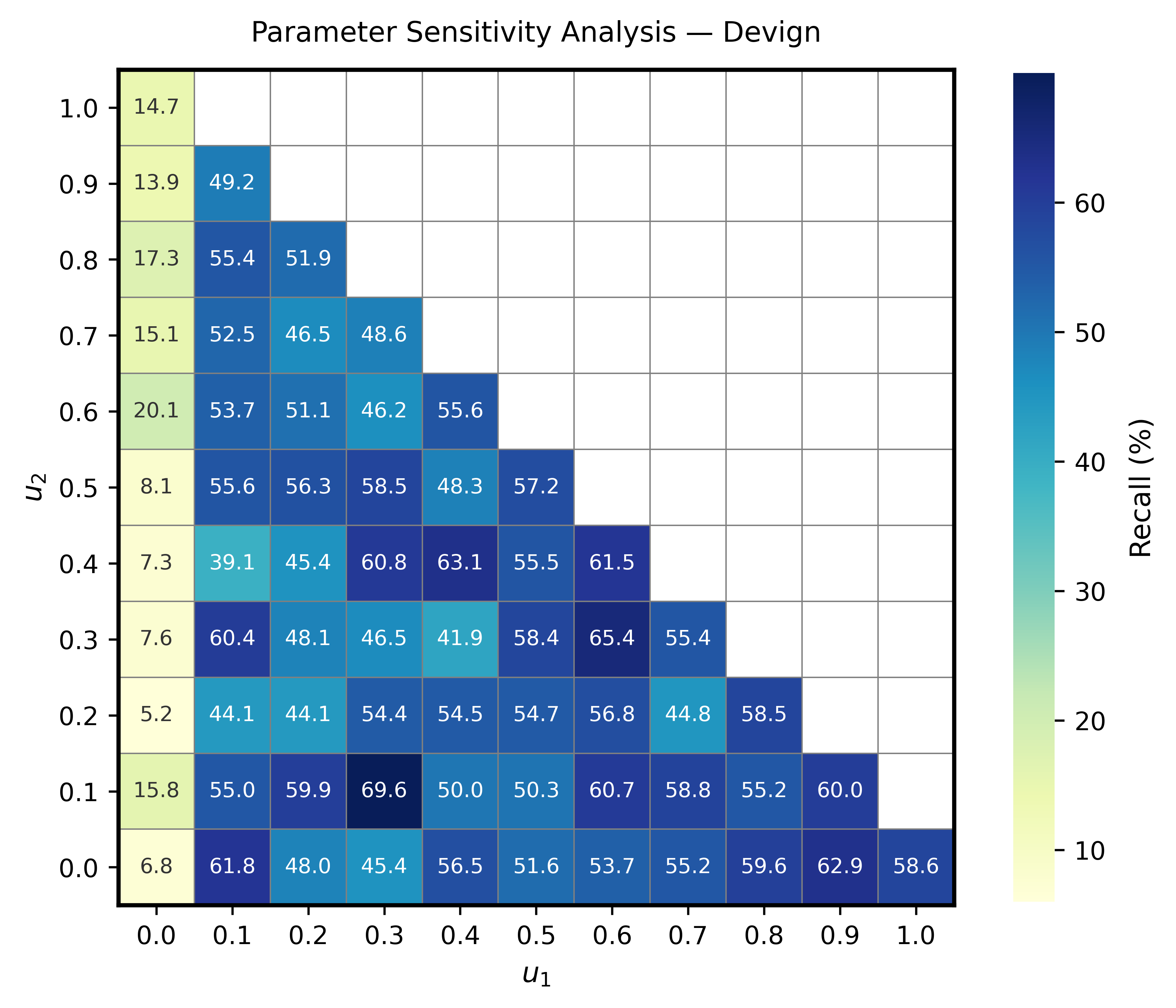}
    \vspace{-0.4cm}
    \label{fig:devign_r}
\end{subfigure}
\hspace{-0.02\textwidth}
\begin{subfigure}{.5\textwidth}
    \centering
    \includegraphics[width=\linewidth,height=0.33\textheight,keepaspectratio]{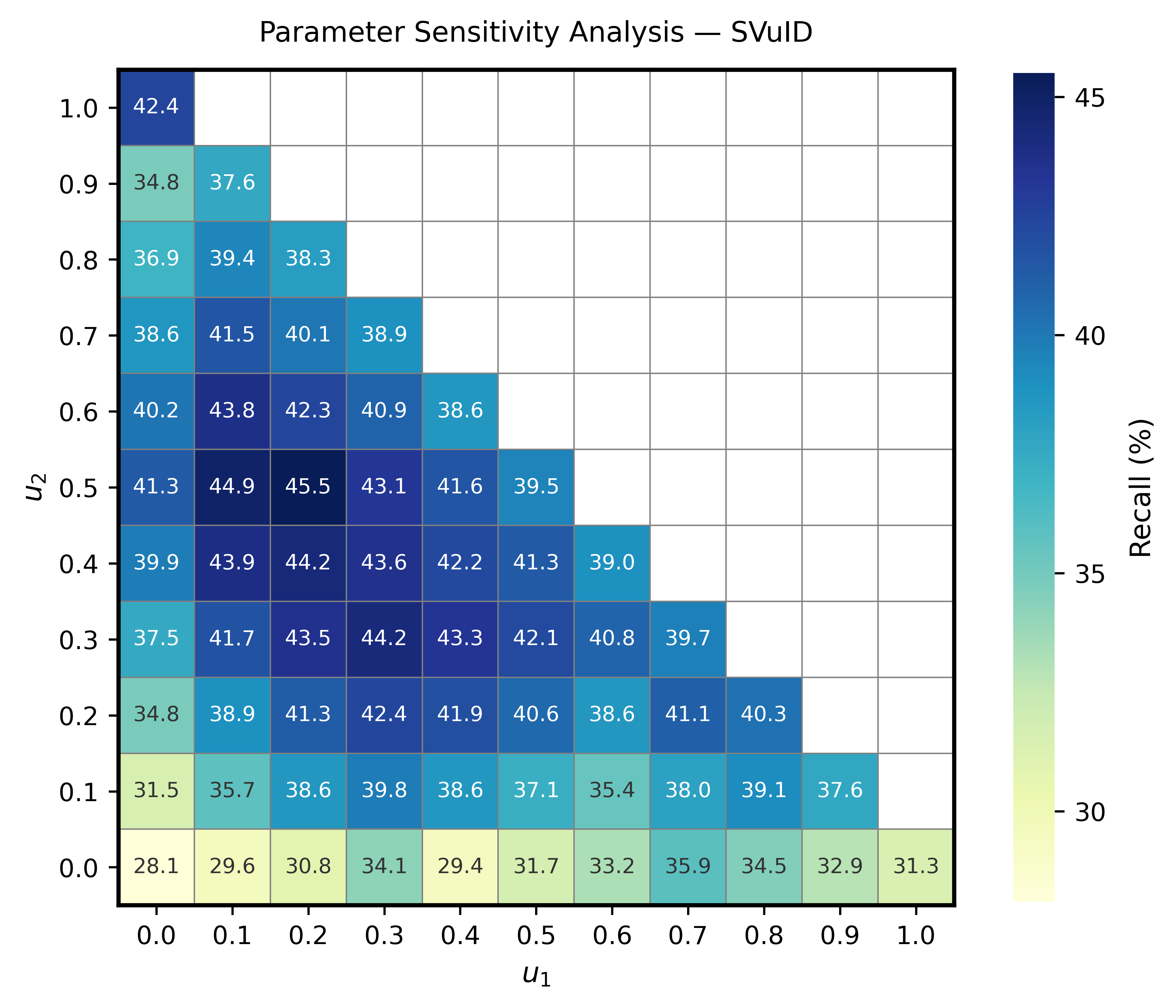}
    \vspace{-0.4cm}
    \label{fig:svuLd_r}
\end{subfigure}

\begin{subfigure}{.5\textwidth}
    \centering
    \includegraphics[width=\linewidth,height=0.33\textheight,keepaspectratio]{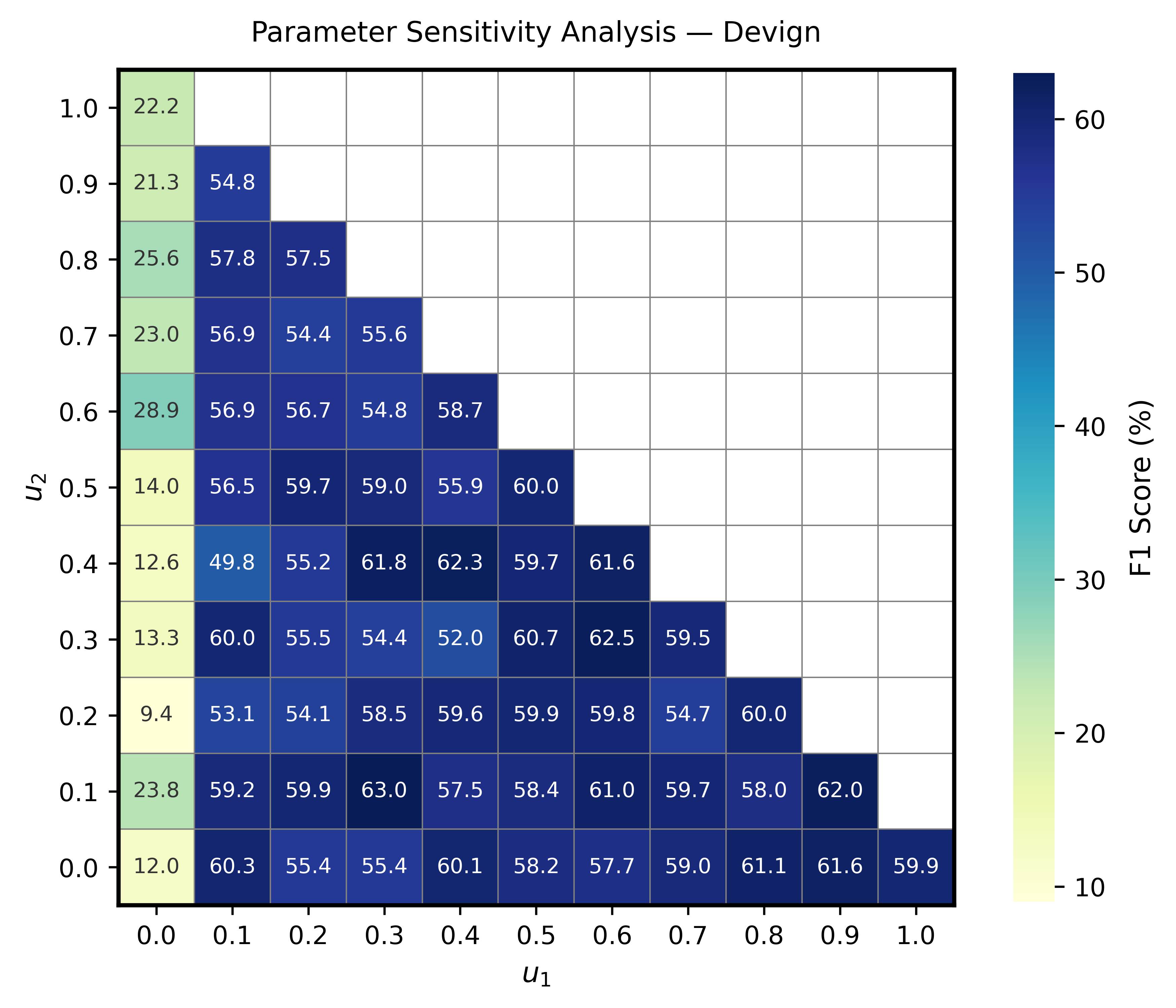}
    \vspace{-0.4cm}
    \label{fig:devign_f1}
\end{subfigure}
\hspace{-0.02\textwidth}
\begin{subfigure}{.5\textwidth}
    \centering
    \includegraphics[width=\linewidth,height=0.33\textheight,keepaspectratio]{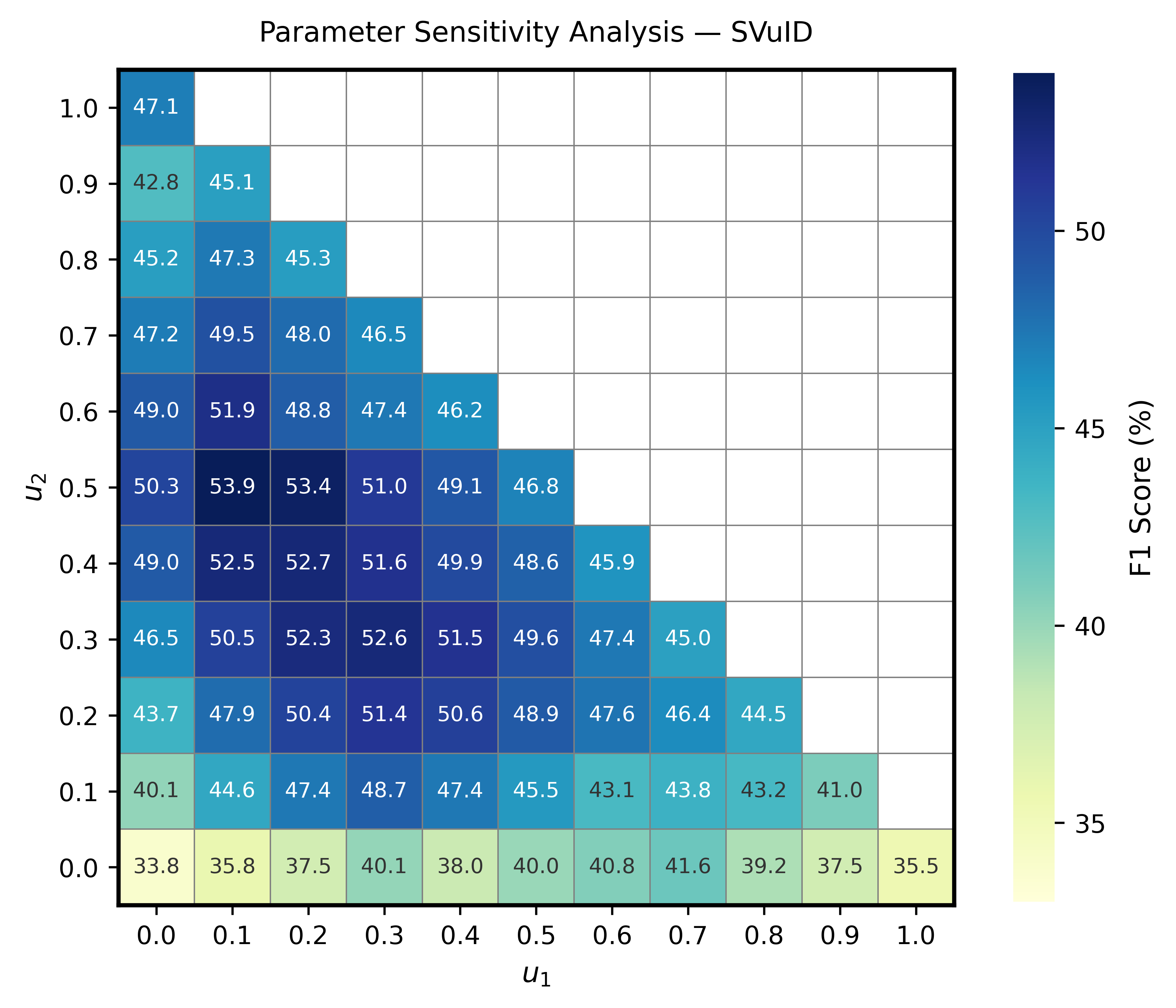}
    \vspace{-0.4cm}
    \label{fig:svuid_f1}
\end{subfigure}
\vspace{-0.2cm}
\caption{Sensitivity Analysis of Parameters $\mu_1$ and $\mu_2$ on Devign and SVulD datasets.}
\label{fig5}
\end{figure*}
}
To address RQ3, we analyze how variations in the SaW parameters $\mu1$ and $\mu2$ influence model performance. As illustrated in Figure~\ref{fig5}, the heatmap analysis across three evaluation metrics shows that, on Devign, 
the model achieves optimal performance when both parameters take moderate values (0.3-0.5). 
Under these conditions, the accuracy, precision, recall, and F1-score reach their respective peaks of 67.6\%, 70.5\%, 69.6\%, and 63.0\%. 
Within this range, all three prediction branches, namely the sequence feature, the structural feature, and the fused feature, are actively utilized, indicating that the SaW strategy effectively balances the relative importance of each branch. 
When examining individual rows or columns of the heatmap, it is evident that increasing either $\mu1$ or $\mu2$ generally leads to an upward trend in F1-score, peaking in the mid-to-high coefficient range. 
Conversely, when $\mu1$ or $\mu2$ approaches extreme values (close to 0 or 1), the performance drops sharply, indicating that relying on a single branch causes significant information loss.
Under an optimal weighting configuration of $\mu1$ = 0.3 and $\mu2$ = 0.1, the model achieves the highest; whereas when $\mu1$ = 0.4 and $\mu2$ = 0.4, it attains the best trade-off between precision and recall.

On SvulD, the model achieves its peak F1-score of 53.9\% and peak precision of 66\% at $\mu1$ = 0.1 and $\mu2$ = 0.5, where a higher weight on the structural branch is required to capture the complex dependency-level vulnerability patterns characteristic of this dataset. Recall reaches its maximum of 45.5\% at $\mu1$ = 0.2 and $\mu2$ = 0.5. Notably, when $\mu2$ = 0 (i.e., the structural branch is entirely excluded), F1-score drops substantially to below 41\%, confirming the critical role of structural features on this dataset. 
These observations suggest that the optimal weighting configuration is dataset-dependent: datasets with explicit token-level patterns favor more balanced branch weights, whereas datasets with subtle semantic distinctions and stronger structural dependencies benefit from a relatively higher structural branch weight.

Overall, the evaluation metrics exhibit nonlinear variations with respect to $\mu1$ and $\mu2$  across both datasets, suggesting that the contributions of different branches to the final prediction are not purely additive but rather interdependent. 
The consistent performance degradation observed at extreme parameter values across both Devign and SVulD further validates the complementary relationship among the sequential, structural, and fused branches under the SaW mechanism, and demonstrates its generalizability across datasets with diverse characteristics.

\subsection{Comparison of Pre-trained Models (RQ4)}
To further explore the impact of different pre-trained encoders on vulnerability detection performance when integrated with GGNN,
we conduct experiments on two representative datasets, Devign and DiverseVul. Devign is an established benchmark widely adopted in prior vulnerability detection studies, whereas DiverseVul provides a more challenging and realistic evaluation setting that better reflects practical detection scenarios.

Within the FusionVul framework, we replace UniXcoder with four widely used pretrained models: CodeBERT ~\cite{23Codebert}, GraphCodeBERT ~\cite{24Graphcodebert}, LongCoder ~\cite{35Longcoder}, and CodeBERT-CPP ~\cite{37Codebertscore}. All other configurations are kept identical to ensure a fair comparison. The experimental results are summarized in Table~\ref{tab9}.

\begin{table}[htbp]
\scriptsize
\centering
\caption{Performance of comparative pre-trained models on Devign and DiverseVul.\label{tab9}}
\begin{tabularx}{\linewidth}{l *{12}{Y}}
\toprule  
&
\multicolumn{3}{c}{\textbf{Devign dataset}} 
& 
\multicolumn{3}{c}{\textbf{DiverseVul dataset}} 
\\
\cmidrule(l){2-4}\cmidrule(l){5-7}
\textbf{Method} & P & R & F1 & P & R & F1 \\
\midrule
CodeBERT  & 64.46 & 43.37 & 51.85 & 33.25 & 14.76 & 20.44 \\
GraphCodeBERT  & 59.77 & 50.44 &  54.71 & 28.95 & 19.34 & 23.19 \\  
CodeBERT-CPP   & 60.81 & \textbf{62.67} & \textbf{61.73} & 30.58 & \textbf{22.74} & \textbf{26.08} \\
LongCoder & 61.96 & 47.47 &  53.75 & 30.79 & 17.95 & 22.68\\  
\textbf{UniXcoder} & \textbf{69.65} & 50.31 & 58.42 
& \textbf{32.10} & 20.64 & 25.12\\
\bottomrule
\end{tabularx}
\end{table}

To address RQ4, we analyze the impact of integrating different pre-trained models into the same FusionVul configuration. 
Table~\ref{tab9} shows that FusionVul achieves the second-best F1 score when UniXcoder is employed as the sequence encoder. Specifically, on the Devign dataset, FusionVul attains an F1 score of 58.42\%, ranking second among all compared pretrained encoders and trailing CodeBERT-CPP (61.73\%). Its Recall is 50.31\%, which is notably lower than that of CodeBERT-CPP (62.67\%) and comparable to GraphCodeBERT (50.44\%), the second-highest Recall among the baselines.
On the DiverseVul dataset, FusionVul achieves an F1 score of 25.12\% and a Recall of 20.64\%, both slightly below those of CodeBERT-CPP (26.08\% and 22.74\%, respectively).
From the observed performance trend, the framework using UniXcoder exhibits consistent behavior across both datasets: it achieves relatively strong Precision but comparatively lower Recall, which in turn constrains further improvements in F1 and leads to a second-best performance that falls behind CodeBERT-CPP on both datasets.

This outcome may be partially attributed to the language-specific enhanced pretraining of CodeBERT-CPP. CodeBERT-CPP is a language-adapted variant of CodeBERT that has undergone additional pretraining on C++ code. Under the same experimental settings, we observe that CodeBERT-CPP consistently achieves higher F1 and Recall than the original CodeBERT. This observation suggests that C++-oriented continued pretraining can enhance the model’s representation capability for vulnerability detection in C++ code, thereby contributing to its superior Recall and overall F1 performance.
However, this advantage is obtained at the cost of additional pre-training data and computational resources.
In contrast, UniXcoder does not require an additional continued pretraining stage, thereby avoiding the associated computational overhead. Despite this, UniXcoder achieves competitive F1 and Recall scores compared with other base encoders under the same experimental setting. 

Overall, these results demonstrate that the choice of pretrained encoder has a significant impact on the performance of multimodal vulnerability detection, and the UniXcoder-GGNN combination reflects a trade-off between detection performance and computational overhead. 

\section{Discussion}\label{sec5}
\subsection{Effects of Weighting Mechanisms and Model Parameters}
FusionVul incorporates multiple parameterized transformations during cross-modal alignment and introduces a weighted aggregation mechanism at the prediction stage. This section clarifies the origin, role, and empirical impact of these weighting rules and associated parameters.

At the structural level, the adjacency weights in Eq. (5) are binary indicators derived directly from the CPG. An entry $A_e[v,u] = 1$ denotes the existence of a syntactic, control-flow, or data-dependence edge; otherwise, it is zero. Message propagation in the GGNN therefore follows program-defined structural relations rather than manually imposed structural preferences, preserving structural consistency in graph modeling.

During cross-modal alignment, the projection matrices and bias terms in Eqs. (8) and (9), including $W_s$, $W_g$, $W_Q$, $W_K$, $W_V$ and $b_s$, $b_g$, are fully learnable parameters optimized end-to-end. These parameters are not manually tuned but automatically adjusted to minimize classification loss. Their role is to align heterogeneous representations into a unified latent space.

At the decision level, the coefficients $\mu1$ and $\mu2$ in the SaW strategy regulate the relative influence of sequential, structural, and fused branches. Sensitivity analysis shows that moderate values outperform extreme configurations, indicating that branch contributions are interdependent. Excessive reliance on a single modality suppresses complementary information and degrades performance.

\subsection{Impact of Dataset Characteristics and Pretrained Encoders}
The effectiveness of these weighting mechanisms is influenced by dataset characteristics and embedding space properties. The four datasets differ in semantic complexity and vulnerability patterns. Devign and Reveal contain relatively explicit lexical cues, where sequence-based encoders can effectively capture discriminative signals. In contrast, SVulD and DiverseVul exhibit subtler semantic distinctions and stronger dependence on control-flow and data-flow relations, requiring deeper structural reasoning. The more stable gains of FusionVul on these semantically complex datasets indicate that multimodal aggregation is particularly advantageous when structural dependencies are critical.

Pretrained encoders further shape embedding characteristics through their pretraining objectives. Language-adapted models such as CodeBERT-CPP tend to improve recall, likely due to better alignment with C++-specific vulnerability patterns, whereas general encoders exhibit more conservative prediction behavior. These differences reflect representation biases induced by distinct pretraining objectives. Importantly, performance variations arise from embedding characteristics rather than architectural dependence, indicating that multimodal aggregation operates consistently across heterogeneous pretrained representations.

\subsection{Interpretation of Empirical Findings}
Taken together, the empirical observations clarify how the proposed mechanisms operate under varying semantic conditions. The stable performance improvement of FusionVul demonstrates that fine-grained multi-modal alignment based on cross-attention is more effective than static concatenation. By selectively amplifying structurally relevant cues and suppressing irrelevant signals, cross-modal attention enhances semantic sufficiency, particularly in datasets where vulnerability patterns are subtle and structurally grounded.

Similarly, the multi-branch aggregation strategy contributes to stability across heterogeneous datasets. Ablation results show that removing either the sequential or structural branch degrades performance, suggesting that vulnerability detection inherently requires complementary lexical and dependency-level evidence. The SaW mechanism preserves modality-specific signals while enabling balanced integration, preventing dominance by any single semantic perspective.

The observed precision-recall trade-offs further support this interpretation. In semantically complex datasets, multimodal constraints reduce false positives by enforcing consistency between lexical and structural cues. Although recall may fluctuate, overall F1 improvements indicate enhanced discrimination in structurally sensitive scenarios. Together, these findings demonstrate that multimodal interaction addresses the intrinsic semantic complexity of real-world vulnerabilities rather than merely extending model architecture.

\subsection{Theoretical and Practical Implications}
The findings yield both theoretical and practical implications for vulnerability detection research and system design. From a theoretical perspective, the results reinforce the view that source code should be modeled jointly as a lexical sequence and a structured semantic graph. Single-modality representations are insufficient to capture the multifaceted nature of program semantics, whereas explicit cross-modal alignment improves representational adequacy. The nonlinear sensitivity of the SaW coefficients further suggests that semantic perspectives are interdependent, indicating that vulnerability detection involves coordinated reasoning across heterogeneous representations. 

From a practical perspective, the results suggest that relying solely on pretrained language models may be inadequate for vulnerabilities involving control-flow or data-flow dependencies. Incorporating structural representations and explicit cross-modal interaction can enhance robustness across diverse datasets. In addition, maintaining separate prediction branches prior to weighted aggregation improves flexibility, allowing practitioners to adjust modality emphasis according to domain characteristics without redesigning the overall architecture. As a modular framework, FusionVul can accommodate alternative pretrained encoders or graph representations, providing adaptability for evolving vulnerability detection environments.

\section{Threats to Validity}\label{sec6}
\subsection{Construct Validity}
Construct validity concerns whether the evaluation metrics and experimental setup accurately measure vulnerability detection capability. In this study, vulnerability detection is conducted at the function level, which may not fully capture statement-level or path-sensitive vulnerabilities. Moreover, the datasets provide binary labels without distinguishing vulnerability severity or exploitability, potentially limiting the granularity of evaluation. Additionally, structural representations are generated using Joern for CPG extraction. Although widely adopted, inaccuracies in static parsing or dependency construction may affect structural completeness and influence downstream modeling.

\subsection{Internal and Statistical Validity}
Internal validity relates to potential confounding factors in experimental design. The SaW coefficients $\mu1$ and $\mu2$ are selected via grid search rather than learned dynamically, which may not guarantee globally optimal weighting across all datasets. Furthermore, the pretrained UniXcoder encoder is frozen during training, and fine-tuning may yield different results.

Class imbalance constitutes another potential threat. As shown in Table~\ref{tab1}, datasets such as Reveal and DiverseVul exhibit skewed distributions. Imbalance may bias prediction behavior and affect precision-recall trade-offs. To maintain comparability with prior work, we preserve original distributions without resampling. Moreover, data leakage is mitigated through dataset-wise deduplication prior to splitting (Section 4.1). While exact duplicates are removed via normalization and hashing, semantic-level similarities cannot be fully excluded.

In addition, from a statistical perspective, experiments use a single random split per dataset. Repeated runs and formal significance testing could further strengthen empirical confidence.

\subsection{External Validity}
External validity concerns generalizability beyond the studied datasets. All experiments are conducted on C/C++ vulnerability datasets. The applicability of FusionVul to other languages or multilingual codebases remains to be validated. Moreover, evaluation is limited to function-level detection, whereas real-world vulnerabilities may span multiple functions or modules. Further investigation is needed to assess scalability and cross-project generalization in industrial environments.

\section{Conclusion}\label{sec7}
This study presents FusionVul, a multimodal vulnerability detection framework that jointly leverages sequential and structural code representations. By integrating a pre-trained UniXcoder with a graph neural network, the framework captures complementary semantic information from linear code sequences and code property graphs. The proposed CAFFNet module further enhances cross-modal interaction through a refinement of attention-based fusion, while the SaW strategy performs weighted aggregation of multi-branch predictions. FusionVul achieves stable performance gains across multiple datasets, indicating that multimodal fusion and multi-branch aggregation enhance the model’s robustness under diverse dataset characteristics.

Future work will extend FusionVul toward large-scale and multilingual code environments to further examine cross-language generalization and scalability.
We also plan to investigate more expressive cross-modal fusion mechanisms, such as adaptive or hierarchical attention architectures, to further improve fusion granularity. In addition, integrating domain knowledge on vulnerability repair and contextual code evolution may enable a closed-loop workflow that supports detection, prioritization, and remediation within a unified pipeline. 
Beyond methodological enhancements, FusionVul could also be incorporated into AI-driven vulnerability management systems as a local analysis component ~\cite{ressi2024ai}. For example, its structured detection outputs may be recorded and aggregated on tamper-evident blockchain ledgers, thereby supporting traceable code auditing and collaborative vulnerability information sharing across organizations.

\section*{Conflicts of Interest}
The authors declare that they have no conflicts of interest or personal relationships that could have appeared to influence the work reported in this paper.

\section*{Author Contribution}
\textbf{Hongyu Yang}: Conceptualization, Methodology, Writing – original draft. 
\textbf{Yaping Zhu}: Methodology, Software, Validation, Writing – original draft. 
\textbf{Jingchuan Luo}: Validation, Writing – review \& editing. 
\textbf{Hiroshi Nomaguchi}: Investigation.
\textbf{Chunhua Su}: Validation.
\textbf{Willy Susilo}: Validation.

\section*{Acknowledgments}
This work was supported by the Civil Aviation Joint Research Fund Project of the National Natural Science Foundation of China [grant number U2433205].

\bibliographystyle{elsarticle-harv}
\bibliography{refs-num}

\end{document}